\documentclass[aps,prb,reprint,groupedaddress,showpacs,amsmath,amssymb,superscriptaddress]{revtex4-2}

\usepackage{graphicx}
\usepackage{caption}
\usepackage{dcolumn}% Align table columns on decimal point
\usepackage{bm}% bold math
\usepackage{float}
\usepackage{placeins}
\usepackage{url}
\usepackage{amsmath,nccmath,amssymb}
\usepackage{subfigure}

\usepackage{calc}

\usepackage{latexsym}
\usepackage{epsf}

\usepackage{epsfig}
\usepackage{notoccite}
\usepackage{euscript}
\usepackage{hyperref}

\usepackage[dvipsnames]{xcolor}
\usepackage{sidecap}

\usepackage[english]{babel}
\usepackage[autostyle, english = american]{csquotes}
\MakeOuterQuote{"}
\usepackage{soul} %For highlighting text with \hl{text}
\usepackage{gensymb}%Allow for symbols such as the degrees symbol to be used
\usepackage{textcomp}%Also allows for symbols such as the degrees symbol to be used, but directly within text.

\begin{document}
\title{Magnetic field tuning of crystal field levels and vibronic states in Spin-ice Ho$_2$Ti$_2$O$_7$ observed in far-infrared reflectometry}

\author{Mykhaylo Ozerov}
%\altaffiliation{These authors contributed equally}
\email{ozerov@magnet.fsu.edu}
\altaffiliation{These authors contributed equally}
%\altaffiliation{National High Magnetic Field Laboratory, FSU, Tallahassee, FL 32310, USA}
\affiliation{National High Magnetic Field Laboratory, FSU, Tallahassee, FL 32310, USA}
 %\altaffiliation{These authors contributed equally}
 
\author{Naween Anand}
%\altaffiliation{These authors contributed equally}
%\altaffiliation{National High Magnetic Field Laboratory, FSU, Tallahassee, FL 32310, USA}
%\thanks{These two authors contributed equally}
\email{anand@magnet.fsu.edu}
\altaffiliation{These authors contributed equally}
%\altaffiliation{These authors contributed equally}
\affiliation{National High Magnetic Field Laboratory, FSU, Tallahassee, FL 32310, USA}

 %{Current address: Materials Science Division, Argonne National Laboratory, Argonne, Illinois 60439, USA}

\author{L. J. van de Burgt}
\affiliation{Department of Chemistry, FSU, Tallahassee, FL 32306, USA}

\author{Zhengguang Lu}
\affiliation{National High Magnetic Field Laboratory, FSU, Tallahassee, FL 32310, USA}
\affiliation{Department of Physics, FSU, Tallahassee, FL 32306, USA}

 \author{Jade Holleman}
  \affiliation{Department of Physics, FSU, Tallahassee, FL 32306, USA}
 \affiliation{National High Magnetic Field Laboratory, FSU, Tallahassee, FL 32310, USA}

 \author{Haidong Zhou}
 \affiliation{National High Magnetic Field Laboratory, FSU, Tallahassee, FL 32310, USA}
\affiliation{University of Tennessee, Knoxville, TN 37996, USA}

\author{Steve McGill}
 \affiliation{National High Magnetic Field Laboratory, FSU, Tallahassee, FL 32310, USA}

\author{Christianne Beekman}
\email{beekman@magnet.fsu.edu}
 \affiliation{Department of Physics, FSU, Tallahassee, FL 32306, USA}
 \affiliation{National High Magnetic Field Laboratory, FSU, Tallahassee, FL 32310, USA}

\date{\today}
\begin{abstract}
Low temperature optical spectroscopy in applied magnetic fields provides clear evidence of magnetoelastic coupling in the spin ice material Ho$_2$Ti$_2$O$_7$. In far-IR reflectometry measurements, we observe field dependent features around 30, 61, 72 and 78~meV, energies corresponding to crystal electronic field (CEF) doublets. The calculations of the crystal-field Hamiltonian model confirm that the observed features in IR spectra are consistent with magnetic-dipole-allowed excitations from the ground state to higher $^5$I$_8$ CEF levels. We present the CEF parameters that best describe our field-dependent IR reflectivity measurements.
Additionally, we identify a weak field-dependent shoulder near one of the CEF doublets. This indicates that this level is split even in zero-field, which we associate with a vibronic bound state. Modeling of the observed splitting shows that the phonon resides at slightly lower energy compared to the CEF level that it couples to, which is in contrast with previously published inelastic neutron measurements. The magnetic field dependence of the vibronic state shows a gradual decoupling of the phonon with the CEF level as it shifts.  This approach should work in pyrochlores and other systems that have magnetic dipole transitions in  the  IR  spectroscopic  range,  which  can  elucidate  the presence  and  the  ability  to  tune  the  nature of vibronic states in a wide variety of materials.
\end{abstract}
\maketitle
\section{Introduction}
In pyrochlore titanates, RE$_2$Ti$_2$O$_7$, the magnetic RE$^{3+}$ ions occupy a lattice of corner-sharing tetrahedra, providing the quintessential framework to study geometrical frustration in three dimensions \cite{Ramirez,Harrisbramwell,harris2}. These systems have been shown to possess a diverse variety of unconventional cooperative magnetic ground states, including spin liquid and spin ice states \cite{GREEDAN,GardnerJ}. The canonical spin ices, Ho$_2$Ti$_2$O$_7$ (HTO) and Dy$_2$Ti$_2$O$_7$, have been studied extensively as they form a two-in/two-out spin configuration on each tetrahedron below $\theta_W \sim$ 2 K \cite{Barry,wiebe,zhou}. This is the result of the very large Ising anisotropy and the long range dipolar interactions that lead to effective ferromagnetic coupling between Ho$^{3+}$ spins \cite{hertog,wiebe,zhou}. Moreover, a large body of recent works has shown that spin ice materials host fractionalized excitations (magnetic monopoles)\cite{gingras2014, Castelnovo, Jaubert}.

The localized spin momentum on the Ho$^{3+}$ is strongly coupled with the 4f orbital momentum and the interaction of the 4f charge cloud with the crystal electronic field from surrounding oxygens leads to the Ising anisotropy found in spin ice materials. 
As pointed out by Ruminy et al., the CEF Hamiltonian is essential to quantify possible quantum corrections to the classical model\cite{Ruminy} in spin ices like HTO. It explains several intriguing phenomena in the rare-earth pyrochlore systems; i.e., the size of the monopole charge, anisotropy of the magnetic moment, interactions with other degrees of freedom such as phonons and spins\cite{Petit,McClarty}, and coupling strength of any transverse spin component. This provides a logical pathway to understand the mechanism of Ising moment reversal allowing monopole dynamics and quantum fluctuations beyond the classical spin-ice limits\cite{Malkin,RauGingras}.
Considerable activity has been devoted to the determination of the crystal field parameters and the corresponding energy-level scheme in HTO \cite{Tomasello,Rosenkranz,Ruminy,Gaudet,Bertin}. Rosenkranz et al. obtained the set of six crystal field parameters based on fitting energies of five CEF transitions measured in an inelastic neutron scattering (INS) experiment. Because the number of the observed CEF transitions was restricted, Bertin et al. suggested a global fitting procedure based on scaling the energy levels available at that time for a variety of rare earth ions within the same pyrochlore RE$_2$Ti$_2$O$_7$ series~\cite{Bertin}. Recently, two detailed experimental INS studies\cite{Gaudet,Ruminy} resolved additional CEF transitions in HTO and included the peak intensities into the fitting. However, there is still some discrepancy in these results. While it has been discussed before that magnetic field can be used to resolve some of the discrepancy\cite{Amelin,Bertin}, our modeling shows that in order to unambiguously determine the CEF parameters the direction of the magnetic field within the local Ho$^{3+}$ ion coordinate frame is an important parameter. 

Magneto-elastic effects are relevant in rare-earth pyrochlores and manifest in terms of modified magnetic, vibrational and electronic properties\cite{Erfanifam}. This effect has been recently reported in HTO through INS measurements\cite{Gaudet}. The measurements showed that the E$_g$ CEF doublet around 60~meV was split due to the coupling with a phonon, evidencing the presence of an entangled phononic crystal field excitation due to strong magneto-elastic coupling\cite{Gaudet}.As we will show, magneto-infrared spectroscopy is a powerful tool that can provide insight into CEF transitions and magneto-elastic effects\cite{Mkaczka, Jandl,Mansouri,Vermette, Lummen} in materials with rather complex CEF schemes, such as HTO. Vibronic states have also been observed recently in other pyrochlore titanates \cite{Amelin,Constable} in the terahertz spectral range. These magneto-optical studies provide a straightforward way to study how CEF-phonon coupling strengths can be varied as a function of applied magnetic field. This work has relevance beyond the pyrochlores. Magneto-IR spectroscopy could provide deeper insights into vibronic states observed in other systems, such as the high T$_C$ superconductor NdBa$_2$Cu$_3$O$_{7-\delta}$. In this superconductor CEF-phonon coupling can be tuned via isotopic substitution of oxygen (to shift the phonon energy) and by applied fields (to shift the CEF levels), providing two ways to tune the nature of the bound state\cite{heyen}.

The main results presented in this paper are, 1) the observation and modeling of magnetic-dipole-allowed transitions between CEF levels and their evolution in applied magnetic field using far-IR reflectivity measurements. We found good agreement between the modeling and our data. 2) finding and modeling the magnetic field dependence of a spectroscopic feature associated with a CEF-phonon coupled (vibronic) state. We modeled the field-evolution of the CEF levels and of the vibron using a phenomenological model. The qualitative comparison between the model and the data allows us to estimate CEF parameters, along with the energy of the phonon that couples to the CEF level.Our work is unique in that it uses IR reflectivity rather than transmission and we show that with IR reflectivity it is possible to observe magnetic dipole allowed transitions between CEF levels and identify the presence of a vibronic state in HTO. There are many works that describe the CEF levels in HTO, but these are mainly neutron studies, where much larger samples are needed and studying magnetic field dependencies (strength and direction) is time consuming and far from trivial. Interestingly, the field dependence has allowed us to clarify that the specific signs of certain CEF parameters can only be distinguished when the field is applied out of the $($110$)$ plane and away from the $<$001$>$ direction in the local Ho$^{3+}$ ion coordinate frame (see Supplementary Materials \cite{suppmat}). Otherwise, this sign issue will go wholly unnoticed. Furthermore, the magnetic field dependence of the vibronic state shows a gradual decoupling of the phonon with the CEF level as it shifts, which has not been reported before. This approach should work in pyrochlores and other systems that have magnetic dipole transitions in the IR spectroscopic range, which can elucidate the presence and the ability to tune the nature of vibronic states in a wide variety of materials.

This paper is organized as follows. We start with details on the experimental (Section \ref{experiment}) setup and important information on the procedure for background correction so that small magnetic field induced changes can be extracted. In section \ref{results} we cover the results, the modeling, and we discuss the significance of our observations. We start with an overview of the data after which we discuss the modeling of our data. We divide the description and discussion of the modeling in subsections, starting with subsection \ref{CEF_zero} about the zero-field transitions and comparing our observations to those reported by others, subsection \ref{CEF_field} about the field dependence of the CEF transitions, and lastly subsection \ref{vibron} about the observation and modeling of the vibronic state in IR-reflectivity. 

\section{Experimental}\label{experiment}
\begin{figure}[t]
\centering
\includegraphics[width= 3in]{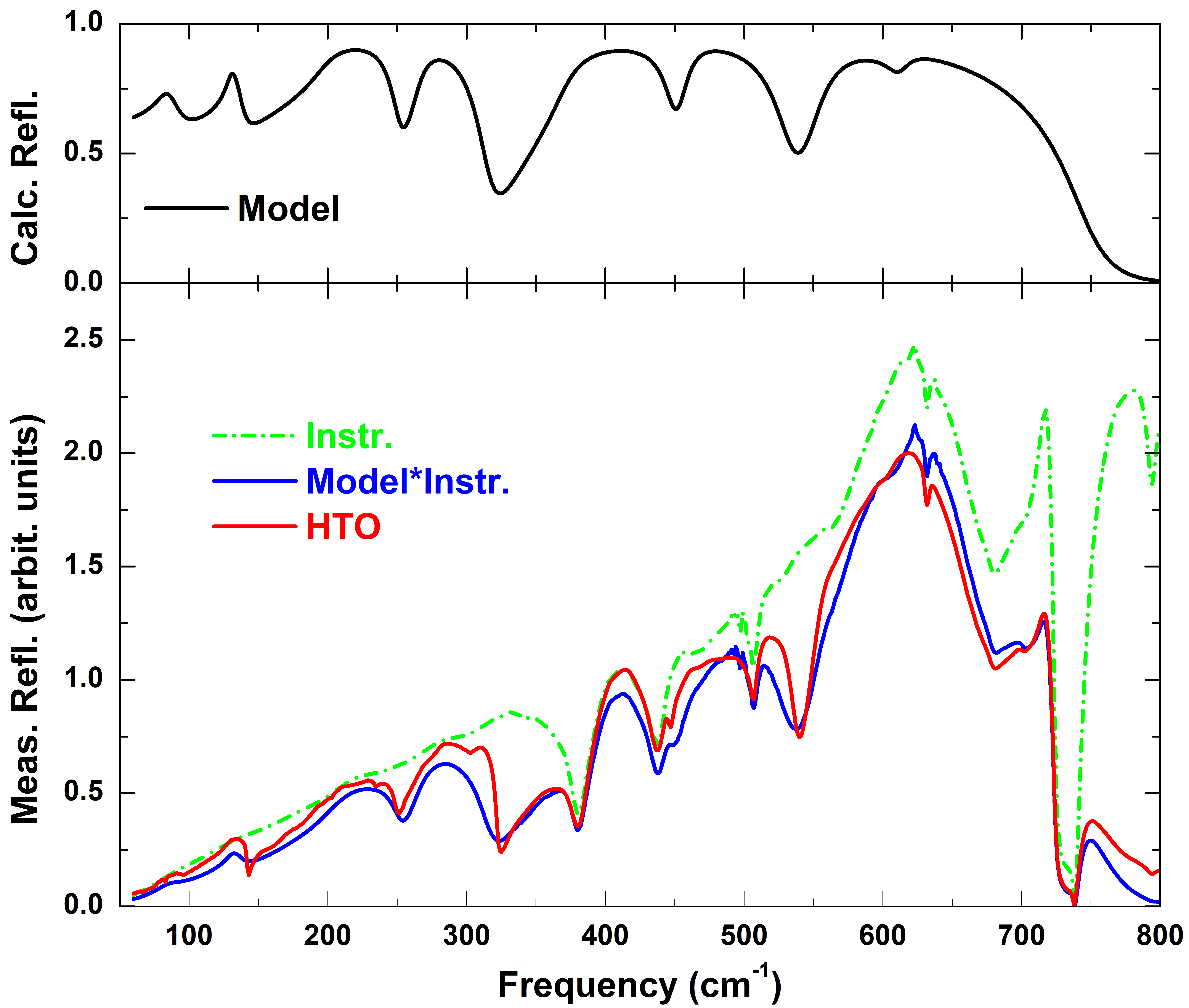}
\caption{\label{IR-zero-field}(Color online) (top) Calculated reflectance of HTO after adjusting the parameters from previously published IR studies on Dy$_2$Ti$_2$O$_7$ single crystals \cite{Bi} to best approximate our measurements (see Table I in the Supplementary Materials \cite{suppmat}). (Bottom) Measured single beam reflected intensity profiles for HTO (red) and a gold reference sample (green). The blue curve shows the multiplication of the model based calculated reflectance (black) and the gold reference (green).}
\end{figure}

The single crystal samples of HTO were grown using the optical floating-zone method. The Ho$_{2}$O$_{3}$ and TiO$_{2}$ powders were mixed in a stoichiometric ratio and then annealed in air at 1450$^{\circ}$C for 40 h before growth in an image furnace. The growth was achieved with a pulling speed of 6 mm/h under 5 atm oxygen pressure. The crystals were oriented by Laue back diffraction. The structural and compositional analyses of these samples were performed previously, confirming the cubic symmetry of crystals with the lattice parameter, in agreement with previously reported values \cite{GardnerJ} (see \cite{Barry} for more details). 

\begin{figure*}[htp!]
\centering
\includegraphics[width= 7in]{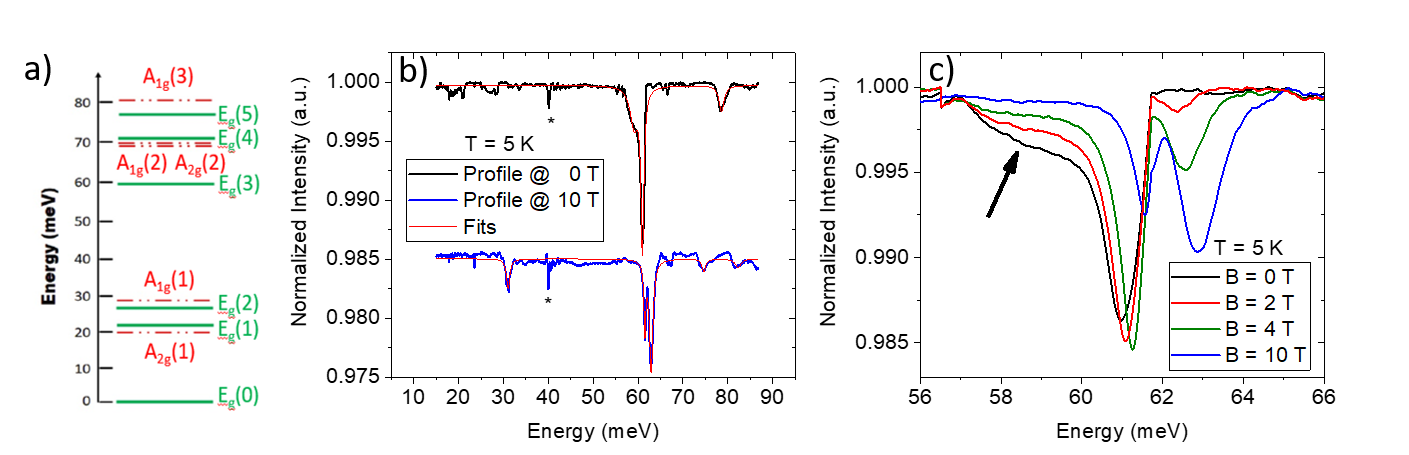}
\caption{\label{HTO}(Color online) a)  Schematic showing the CEF energy levels\cite{Rosenkranz} for the Ho$^{3+}$ ion in Ho$_2$Ti$_2$O$_7$. The green solid levels are the E$_g$ doublets and the red dashed levels are the A$_{1g}$ and A$_{2g}$ singlets, respectively. 
b)Normalized reflection spectra measured for magnetic fields of B = 0~T (black line) and B = 10~T (blue line, with added offset). The red line is a cumulative fit of three lorentz peaks at 59.11, 60.96, and 78.60~meV, respectively. An instrumental artifact is marked by the asterisk. c) Normalized spectra at several magnetic fields in the vicinity of 60~meV. The arrow shows the shoulder slightly below the strongest CEF transition.}
\end{figure*}

The magneto infrared spectroscopy was performed at the National High Magnetic Field Laboratory employing a 17~T vertical-bore superconducting magnet coupled with Fourier-transform infrared spectrometer Bruker Vertex 80v. The collimated IR radiation was propagated from the spectrometer to the top of the magnet inside the evacuated ($\approx$ 4~mBar) optical beamline and then focused to the brass lightpipe, used to guide the IR radiation down to the sample space of the magnet. The parabolic 90\degree degree mirror focused the IR radiation on the sample with  $\approx$ 30\degree degree incident angle, while a second confocal mirror collected the reflected IR radiation inside the twin lightpipe with the Si composite bolometer at the end. The reflective surface of the sample was oriented parallel (Voigt geometry) to the magnetic field applied along $[$001$]$ crystallographic direction. The reflection spectra were measured in the spectral range between 50--800 cm$^{-1}$ with instrumental resolution of 0.3 cm$^{-1}$. Both sample and detector were cooled by low-pressure helium gas to a temperature of 5~K. The experimental information about Raman is discussed in the Supplementary Materials \cite{suppmat}. 

The signal-to-noise ratio of the magneto-infrared data was improved by averaging over three spectra collected at every field point. Then, the single beam spectrum at each magnetic field was divided by a reference spectrum to remove a strong non-magnetic background signal and thereby to disclose the tiny field-dependent features. The spectra measured at all magnetic fields were combined into the 2D matrix, with rows and columns corresponding to energy and magnetic field points, respectively. The reference spectrum is created by taking the highest value of the intensity at each column (i.e. at each frequency point) of this 2D spectrum. The normalization on such statistically created baseline keeps the relative reflectance spectrum below 100\% and quantifies the field-induced changes in the reflection signal. The statistical approach for the background correction is frequently used to process transmission data and causes the field-dependent feature to possess a peak line shape, instead of the peak-derivative shape, intrinsic to the normalization on the zero-field spectrum. For instance, Amelin et al. \cite{Amelin} employed the approach in their THz transmission study of CEF excitations in another pyrochlore compound, Tb$_2$Ti$_2$O$_7$. Interestingly, this approach also works very well for the analysis of the reflection data presented in this paper.

\section{Results and Discussion}\label{results}
In Fig. \ref{IR-zero-field} we show single beam reflected intensities collected for HTO (red curve) and for a gold standard (green dashed curve), taken under similar instrumental conditions. However, the direct ratio method is not a viable option for extracting the absolute reflectance for HTO due to inevitable small mismatches in the optical path and hence the appearance of reflectance values above 100$\%$. Instead, we optimize the previously published Lorentzian parameters for a Dy$_2$Ti$_2$O$_7$ single crystal \cite{Bi} resulting in a calculated reflectance spectrum (black curve, top panel in Fig. \ref{IR-zero-field}). We scale the gold standard intensity profile with this calculated reflectance and superimpose the result over the measured reflected intensity profile for HTO (blue curve, bottom panel). The comparison between the experimental intensity profile and the model based calculated reflection intensity profile shows good agreement in the entire frequency range of our interest. While the resonance frequencies barely differ from Dy$_2$Ti$_2$O$_7$ vibrational spectra, the linewidth and oscillator strength for a few phonons show slight variations for HTO. We provide a table in the Supplementary Materials \cite{suppmat} listing all the transverse and longitudinal modes used to calculate the reflectance curve.

The relative changes of the IR spectrum induced by applied magnetic field are shown in the Fig.\ref{HTO}b. The magnetic field in our study is applied along the [001] crystallographic direction, which provides the largest net magnetic moment projection and the same CEF level splitting for all four crystallographically different Ho$^{3+}$ sites \cite{suppmat}. The significant field-induced responses are found at energies of 30 (visible in 10~T spectrum), 61, 72 (visible in 10~T spectrum), and 78~meV. These energies are in line with previous INS studies\cite{Rosenkranz,Ruminy,Gaudet} and we can associate these features with CEF excitations from the ground state doublet E$_g$(0) to higher-energy states, shown in Fig. \ref{HTO}a. 

The largest change in the IR reflection spectra is located in the vicinity of 60 meV, that corresponds to the strongest peak in the INS intensity spectrum measured in zero magnetic field and at low temperature~\cite{Gaudet}. This INS peak appears to have a satellite peak with smaller intensity at the low-energy side. Such splitting is explained by phonon and E$_g$(3) CEF level hybridization~\cite{Gaudet}.  Concurrently, the strongest feature of the normalized IR reflection is also different to other peak-like features due to the presence of a shoulder on the low-energy side (see Fig.\ref{HTO}c). Moreover, the shoulder disappears quickly in applied fields of a few Tesla and we will discuss such behavior in more detail in the text below.

\begin{figure*}[h]%\vspace{-1in}
\includegraphics[ width=0.8\textwidth]{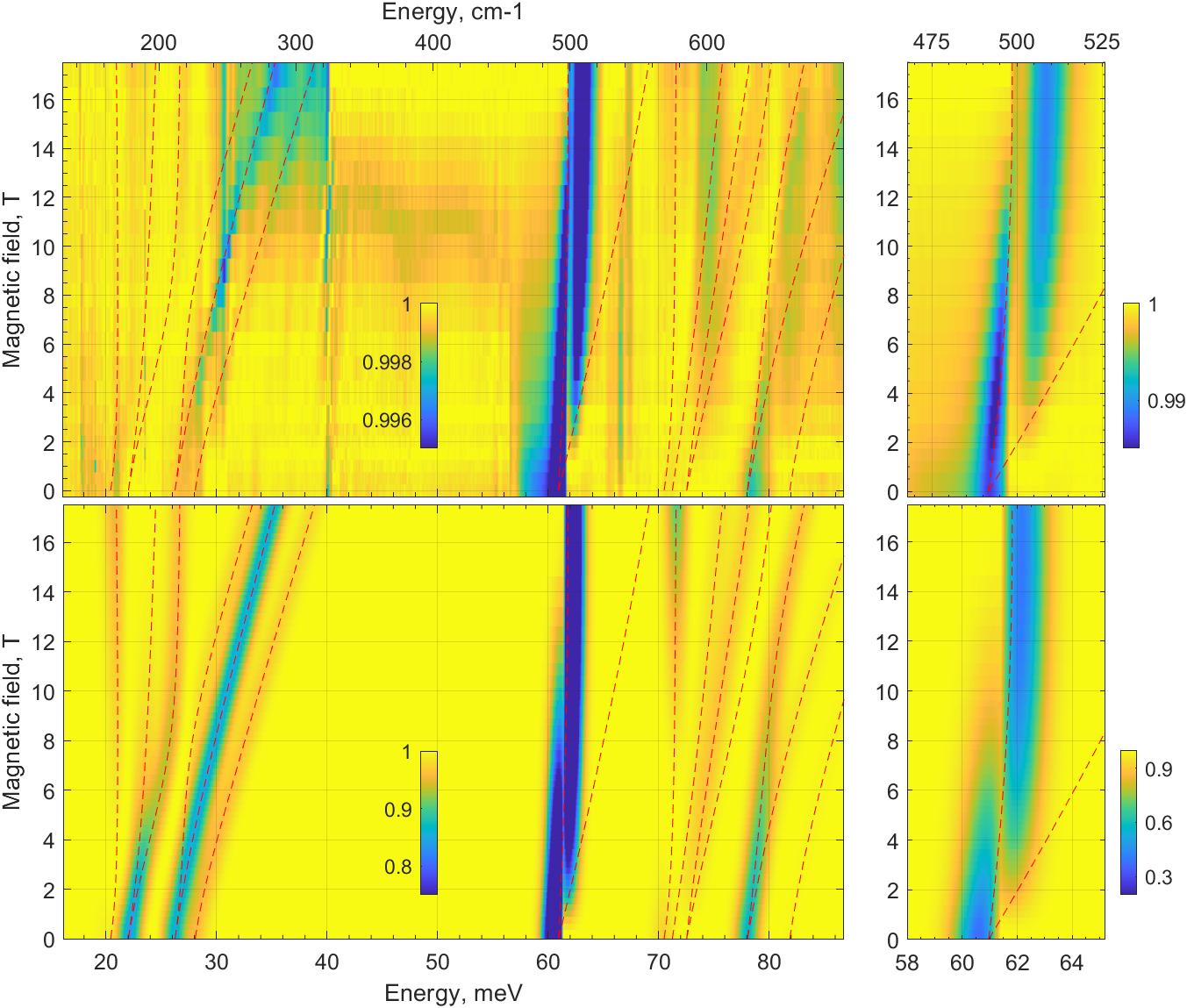}
\caption{\label{exp} (color online)  (Top) The normalized far-IR reflection spectrum as a function of the applied magnetic field. (Bottom) The simulation of the transition intensity between ground and high-energy CEF levels. The field dependencies of the transition energy are plotted with red dashed lines. The calculations used CEF parameters presented in the bottom row of Table \ref{t1}.  Both experimental and calculated spectra were normalized to the reference spectrum obtained from the statistical approach described in the text, respectively. The panels on the right show field-evolution of the strongest CEF transition in the vicinity of 61 meV.} 
\end{figure*}

Interestingly, we can roughly compare the strength of the electric and magnetic dipole excitations in HTO. While the IR active phonons induce changes of $>$ 25\% to the ideal 100 \% reflectivity (Fig.\ref{IR-zero-field}), the intensity changes of CEF transitions are about 0.2\%-1\% of the magnitude of the normalized reflectivity (Fig.\ref{HTO}b). Owing to the high sensitivity of the magneto-infrared spectroscopy technique, the weak magnetic dipole transitions can be still detected in the broadband spectral range. This enables us to investigate the evolution of CEF levels with applied magnetic field in a straightforward way using far-IR reflection measurements, in addition to the magnetoinfrared transmission studies, which are restricted in the spectral range by the transparency windows of the sample. The complete 2D (vs field and energy) spectrum of normalized reflection is presented in the Fig.\ref{exp}, top panel.

\onecolumngrid

\vspace{0.5 cm}
\twocolumngrid
\FloatBarrier

\subsection{The CEF Hamiltonian, an overview of zero-field CEF parameters.} \label{CEF_zero}
 The pyrochlore compound Ho$_2$Ti$_2$O$_7$ belongs to the cubic space group Fd$\bar{3}$m, where the Ho$^{3+}$ ions are sitting on sites of antiprismatic trigonal symmetry D3d~\cite{Tomasello}.
Hence, the crystal field Hamiltonian in applied magnetic field can be conveniently expressed as the sum of seven terms \cite{Tomasello,Bertin,Gaudet} as

\begin{equation} 
\begin{aligned}
\mathcal{H_{CEF}} = B^{0}_{2}\widehat{O}^{0}_{2}+
B^{0}_{4}\widehat{O}^{0}_{4}+
B^{3}_{4}\widehat{O}^{3}_{4}+B^{0}_{6}\widehat{O}^{0}_{6}+\\
B^{3}_{6}\widehat{O}^{3}_{6}+
B^{6}_{6}\widehat{O}^{6}_{6}+ g_L\widehat{J}H
\label{eq1}
\end{aligned}
\end{equation}

where $\widehat{O}^{q}_{k}$ are the extended Stevens operators and $B^{q}_{k}$ are the associated coefficients. The last term is the Zeeman energy defined by the Lande g-factor $g_L=5/4$, angular momentum operator $\widehat{J}$ ($|J| = 8$), and the magnetic field $H$, applied along $<$111$>$ axis in the local Ho$^{3+}$ coordinate frame ([001] direction in the lab frame)~\cite{suppmat}.
We solved this problem and calculated the intensity of magnetic-dipole allowed transitions at $T=0$ K using the Easyspin package \cite{Easyspin,FD-easyspin} in Matlab.

\begin{table*}[htp]
  \caption{Summary of CEF coefficients in [meV] taken from Ref.\cite{Ruminy,Rosenkranz,Gaudet,Bertin,Tomasello} and determined in this work.}
    \label{t1}
		\resizebox{0.8\textwidth}{!}{

\begin{tabular}{|l|l|l|l|l|l|l|} 
		\hline
                        &$B^{0}_{2}$ & $B^{0}_{4}$ & $B^{3}_{4} $& $B^{0}_{6}$ & $B^{3}_{6}$ & $B^{6}_{6}$\\ \hline
			Ref\cite{Tomasello}    &$-7.6e^{-2} $        &$ -1.1e^{-3}$        &$ 8.2 e^{-3} $&$ -7.0e^{-6} $&$ -1.0e^{-4} $&$ -1.3e^{-4}$ \\ \hline   
			 Ref\cite{Bertin} &$-6.8e^{-2}$         &$ -1.13e^{-3}$       &$ -1.01e^{-2} $&$ -7.4e^{-6} $&$ 1.23e^{-4} $&$ -1.01e^{-4}$ \\ \hline
						Ref\cite{Gaudet,Freeman}    &$-8.181e^{-2} $        &$ -1.153e^{-3}$        &$ -8.175 e^{-3} $&$ -6.87e^{-6} $&$ 1.021e^{-4} $&$ -1.309e^{-4}$ \\ \hline
						 LS-coupling\cite{Ruminy}&$-7.811 e^{-2}$ &$ -1.17e^{-3}$       &$ -8.03e^{-3} $&$ -7.07^{-6} $&$ 1.03e^{-4} $&$ -   1.33e^{-4}$ \\ \hline
			 This work &$-7.558 e^{-2}$ &$ -1.156e^{-3}$       &$ -8.685e^{-3} $&$ -7.3e^{-6} $&$ 1.060e^{-4} $&$ -   1.264e^{-4}$ \\ \hline
			\end{tabular}
		}
						
\end{table*}

There are several prior studies that report values for the CEF parameters $B^{q}_{k}$. Rosenkranz et al. \cite{Rosenkranz} report the parameters determined from transition energies observed in INS experiments. Recently, two detailed experimental INS studies \cite{Ruminy,Gaudet} clearly resolved CEF transitions in HTO and, furthermore, allowed to include their relative intensities into the fit constraints.  
Although Gaudet et al. \cite{Gaudet} observed the most intensive peak at 61 meV, it was concluded that the corresponding CEF transition was at 58.9 meV. Such red shift was attributed to the hybridization of E$_g$ level with the silent phonon via vibronic coupling. Using Stevens re-normalization procedure \cite{Steve} we reproduce the CEF parameters from Bertin et al. \cite{Bertin}, Rosenkranz et al. \cite{Rosenkranz} (same as Tomasello et al. \cite{Tomasello}), Gaudet et al. \cite{Gaudet}, and Ruminy et al. ($LS$-coupling scheme \cite{Ruminy}) in Table \ref{t1}. The main differences between all of these sets are i) the sign of the coefficients $B^{3}_{4} $ and $B^{3}_{6}$, and ii) a relatively large spread ($\sim$ 20$\%$) in most of the $B^{q}_{k}$ values also becomes apparent.

\begin{figure}[htb]
\includegraphics[width=0.37\textwidth]{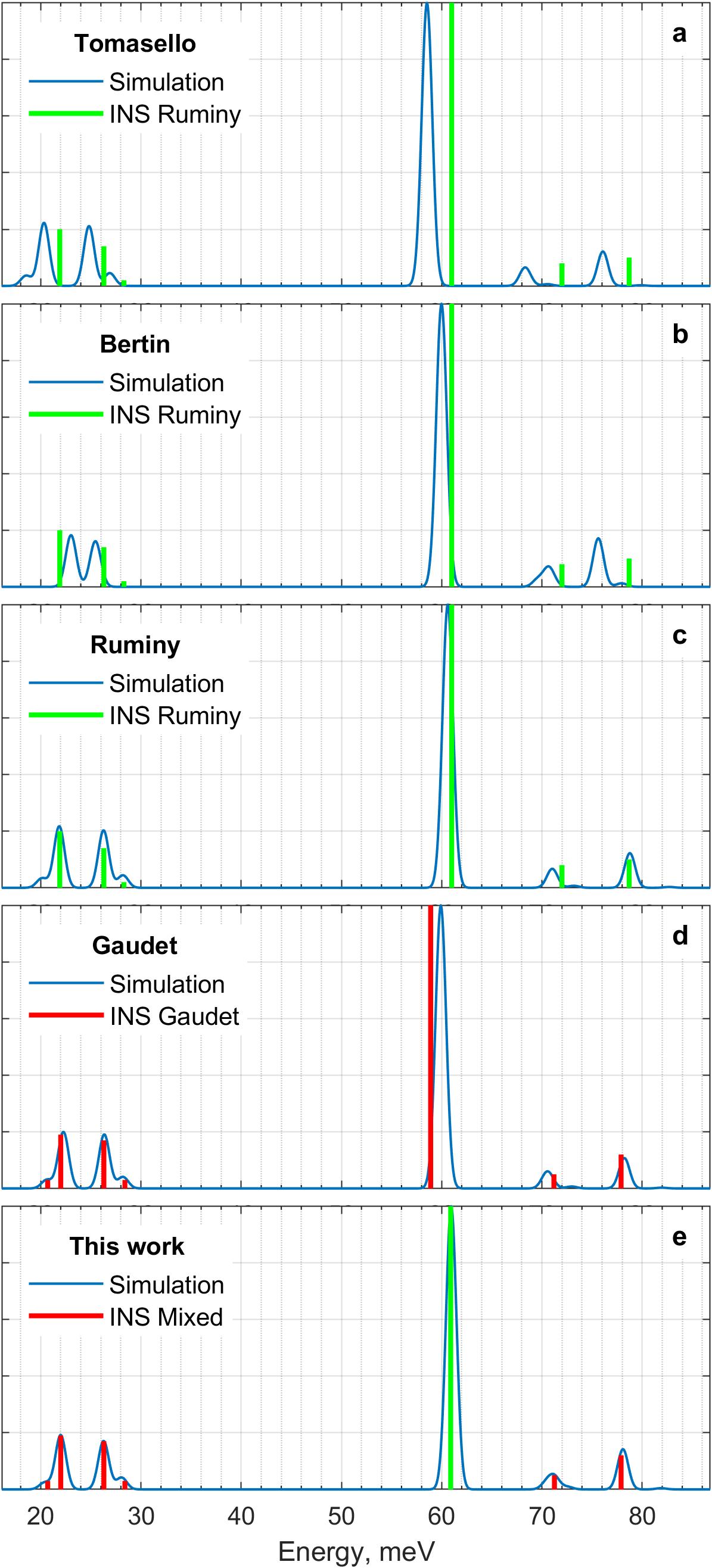}
\caption{\label{INS_compare} (color online)  Zero-field CEF excitation spectrum observed by INS studies (green bars\cite{Ruminy} and red bars \cite{Gaudet}). 
The blue lines are simulations of CEF transitions using Hamiltonian in eq. \ref{eq1} and Stevens coefficients shown in Table \ref{t1}. } 
\vspace{+0.3cm}
\end{figure}

 The CEF excitation spectra were simulated for zero magnetic field for each of the parameter sets shown in Table \ref{t1}. are shown in the Fig. \ref{INS_compare}, along with the intensities of the experimentally observed INS peaks (Table \ref{t2}). The simulation deviates from the observed values (see Fig.\ref{INS_compare}), which is related to complexity of multiple-parameter fitting as well as to the discrepancy in the peak assignment. The simulated CEF excitation spectrum using CEF parameters from Tomasello et al~\cite{Tomasello,Rosenkranz} (Fig.\ref{INS_compare}a)  does not agree with the INS results from Ruminy et al. \cite{Ruminy}. Similar deviations are found when comparing the CEF excitation spectrum using Bertin et al. \cite{Bertin} CEF parameters (Fig.\ref{INS_compare}b). In panel c) of the same figure, the simulated spectrum using the LS-coupling scheme CEF parameters \cite{Ruminy} is compared to the INS observations and overall the agreement is good. 
 Lastly, Ref.\cite{Gaudet} (Fig.\ref{INS_compare}d), the position of the strongest peak observed in the INS experimental spectrum is different compared to the calculated nominal energy of the E$_g$(0) $\rightarrow$ E$_g$(3) transition based on the presented CEF parameter values. 

Ideally, we would simulate a similar zero-field spectrum solely based on our IR spectroscopy results. However, given that we only observe two of the expected seven transitions in zero applied field, this would not produce reliable CEF parameter values. Hence, we adopt the following optimization approach: 
we use reported INS data (Table \ref{t2}) that most closely matches our IR spectroscopy data and fit both energies and intensities of INS peaks using CEF Hamiltonian eq.\ref{eq1} in zero field (Fig.\ref{INS_compare}e). The obtained CEF parameter values that best reproduce those transition energies and intensities are further used for simulations with applied fields without any other adjustments.

  Note, these optimized CEF parameters are considerably different from previously reported values for three main reasons, 1) the spread in the observed transition energy for E$_g$(3), for optimization we selected 61~meV for this transition, in line with our IR spectroscopy data and most closely matching the transition energy reported by Ruminy et al. \cite{Ruminy} (see Table \ref{t2}); 2) The simulations based on previously reported CEF parameter values show a mismatch when compared to the reported observed CEF transition energies; 3) The reported CEF parameter values consistently underestimate the observed transition energy that occurs around 70 ~meV, which is likely a sum of the E$_g$(4) and A$_{2g}$ transitions. To give a sense of how CEF parameters affect the transition energies, we show the shift of each of the CEF levels (about 1~meV) for a 5$\%$ variation of each of the CEF parameters, see Supplementary Materials \cite{suppmat}.

\begin{table}[htp]
  \begin{center}
    \caption{The experimentally observed and calculated CEF energies in Ho$_2$Ti$_2$O$_7$ [meV] at zero magnetic field.}
    \label{t2}
    \begin{tabular}{|l|l|l|l|l|l|}
		\hline
                             CEF& E$_{obs}$\cite{Gaudet} & I$_{obs}$\cite{Gaudet}  & E$_{obs}$\cite{Ruminy} & I$_{obs}$\cite{Ruminy}   & E$_{cal}$  \\ \hline
														$E_g$   &   0  & 0    & 0     & 0      &0       \\ \hline
														$A_{2g}$& 20.7 & 0.03 & $-$   & $-$   &20.42   \\ \hline
														$E_g$   & 22   & 0.19 & 21.9  & 0.2   &22.024   \\ \hline
														$E_g$   & 26.3 & 0.17 & 26.3  & 0.14  &26.24    \\ \hline
														$A_{1g}$& 28.4 & 0.03 & 28.3  & 0.02  &28.07   \\ \hline
														$E_g$   & 58.9 & 1    & 61.0  & 1     &60.96 \\ \hline
														$A_{1g}$& $-$  &$-$   & $-$   & $-$   &70.51  \\ \hline
														$A_{2g}$& 71.2 & 0.05 & $-$   & $-$   &71.26  \\ \hline
														$E_g$   & $-$  & $-$  & 72(1) & 0.08  &72.55 \\ \hline
														$E_g$   & 77.9 & 0.12 & 78.7  & 0.1   &78.05 \\ \hline
														$A_{1g}$& $-$  & $-$  & $-$   & $-$   &81.89   \\ \hline
													
    \end{tabular}
  \end{center}
\end{table}

\subsection{Modeling of field-dependent IR reflectivity data}\label{CEF_field}
We use the optimized zero-field $B^{q}_{k}$ coefficients (last row in Table \ref{t1} ) to model the field-dependence of the CEF transitions and to compare it with experimental observations. The top panel of Fig.\ref{exp} shows the color map of measured intensities as determined from magneto-IR reflection spectroscopy, along with the calculated Zeeman splitting of the CEF levels (red lines). The data exhibits a very good agreement with the shifts in the calculated CEF transitions for E$_g$(2) (at $\approx$26~meV in zero field), E$_g$(3) (at $\approx$61~meV in zero field),  and E$_g$(5) (at $\approx$78~meV in zero field). This agreement is striking, as the simulated intensity is appropriate for transmission experiments, while our IR spectra are measured in a reflection geometry. Our measurements do not show a clear transition associated with the two lowest energy CEF levels around 20~meV. This is due to low sensitivity of our measurement in this energy range. Furthermore, at low field, the IR transitions will be prone to thermal broadening, making them harder to observe and model. It appears that a field of 5~T or greater is needed to resolve some weaker features in the IR spectra.

It is worth to note, that zero-field INS spectra can be equally fitted with two sets of the CEF coefficients, with the difference being the sign of the $B^{3}_{4} $ and $B^{3}_{6}$ coefficients. The sign of these coefficients does alter the magnetic field induced splitting of some CEF levels, but how the transition energies are affected depends on the direction of the magnetic field within the crystal field frame~\cite{suppmat}. With the field applied in the xy-plane~\cite{Tomasello} or along the z-axis in the Ho$^{3+}$ local frame, the resultant transition energies appear invariant upon a sign change of the $B^{3}_{4} $ and $B^{3}_{6}$ coefficients. Only if the field is applied away from these directions, like along on of the $<$111$>$ axes (which is the case in our experiment), the transitions energy become sensitive to this sign change (see Supplementary Materials \cite{suppmat}). In agreement with previous report for Tb$_2$Ti$_2$O$_7$ ~\cite{Amelin}, we experimentally distinguished signs of the CEF parameters in HTO using applied magnetic fields (Fig.S13 in Supplementary Materials \cite{suppmat}). Furthermore, if we use the previously reported Stevens coefficients (see Table I) and calculate the field dependence of the CEF levels, we find the agreement with our IR data to be far less \cite{suppmat}. This shows that magneto-IR is effective in characterizing the field dependence of CEF levels and that some Stevens operators can be determined with a greater degree of accuracy, at least compared to zero field measurements using other probes.

To further compare our optimized CEF parameters to previously reported values, we determined and tabulated the wavefunctions for each of the $m_J$ values of the ground state multiplet for all parameters in Table \ref{t1} (see Supplementary Materials \cite{suppmat}). Quantum corrections to the classical model \cite{RauGingras,Ruminy} stem from spectral content of sub-leading components of the wave function. As expected, we find the spectral content to be predominantly $|\pm8>$ with the sub-leading components of the wave function comparable to those presented by others \cite{Tomasello,Ruminy,Bertin}. 

\subsection{Observation and modeling of vibronic states.}\label{vibron}
While the equation \ref{eq1} satisfactorily describes the splitting of  the CEF levels in applied magnetic fields, this model does not explain the appearance of the field-dependent shoulder observed on the low energy side of the 61~meV transition. This shoulder clearly indicates that this CEF level is split even in zero field. This observation is consistent with previously reported INS measurements \cite{Gaudet}, which reported evidence of overlapped vibrational and electronic degrees of freedom, resulting in a vibronic bound state around the same doublet transition energy. Density functional theory (DFT) calculations by others\cite{KUMAR,Ruminy,Kushwaha} have reported the presence of an optically-silent phonon mode of E$_u$ symmetry in the close vicinity of the E$_g$(3) doublet transition. In the following, we will model this behavior and extract an energy for the phonon that results in the observed CEF-phonon hybridization.

To model the shoulder in the vicinity of 61 meV in the IR spectra, we are solving the following Hamiltonian eq.\ref{eq2}, similar to the previous reports \cite{Gaudet}.
\begin{equation} \label{eq2}
\begin{aligned}
\mathcal{H}_{tot} = \mathcal{H_{CEF}} + \hbar\omega(\hat{a}^\dagger\hat{a}+1/2)-\sum_{q=-2}^{+2}g_{q}(\hat{a}^\dagger+\hat{a})\widehat{B}_{2}^q\widehat{O}_2^q
\end{aligned}
\end{equation}
Here, $H_{CEF}$ is the crystal field Hamiltonian with the Zeeman term (eq.\ref{eq1}). The operators $\hat{a}^{\dagger}$ and $\hat{a}$ correspond to the creation and annihilation of a phonon. The last term represents the vibronic Hamiltonian \cite{Amelin,Gaudet} with a phenomenological coupling constants $g_{q}$ and quadrupolar operators $\widehat{O}_2^q$.   
 The coupling constants were taken $g_{\pm2}=g_{0}$ and $g_{\pm1}=2g_{0}$ to provide the same weight for the angular momentum operators as in the Ref~\cite{Gaudet}. The IR response is proportional to the transition matrix element of the magnetic-dipole operator, which we calculated between the lowest-energy E$_g$(0) doublet state and four states, resulting from the coupling of the E$_g$(3) doublet and E$_u$ phonon.
 
\begin{figure}[htp]
\centering
\includegraphics[width=0.48\textwidth]{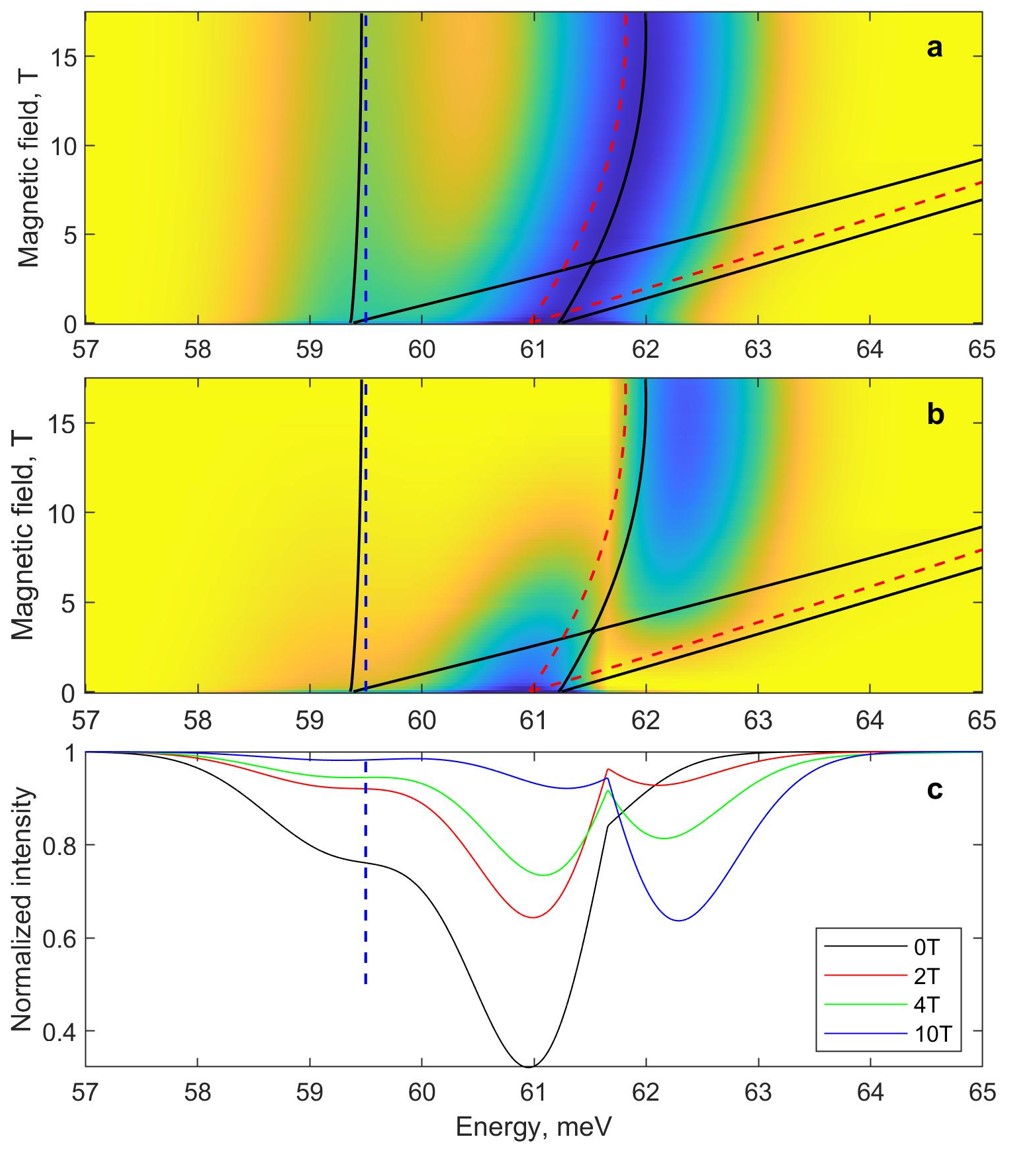}\caption{\label{CEF-phonon} (color online)  a) Illustration of the excitation spectrum from ground to hybridized states as a function of the magnetic field.  The intensity of the transitions is calculated for $T=0$~K, $\hbar\omega=59.5$~meV, coupling constant $ g_0=0.016$  and Lorentizan linewidth of 1.6~meV. The dashed red and blue lines correspond to the  $E_g(3)$ CEF doublet and phonon mode, respectively. The black solid lines show energies of hybridized states. b) The same excitation spectrum but now normalized by a reference spectrum, calculated in the same way applied for the Fig. \ref{exp}. c) Profiles taken at various fields based on the middle panel.}  
\end{figure}

Fig. \ref{CEF-phonon} a) shows the intensity associated with the E$_g$(3) CEF excitation in the presence of phonon-CEF hybridization, with the phonon energy $\hbar\omega=59.5$~meV indicated by the dashed blue line. 

The red dashed lines are the same as in Fig. \ref{exp}. The colormap clearly indicates the presence of a much broader feature around 61~meV compared to the simulated IR intensity using $H_{CEF}$ alone. To compare to our measured data we apply the same normalization routine to the calculated data as before, which results in the color map in Fig.~\ref{CEF-phonon} (b). Profiles taken at B = 0, 2, 4 and 10 T result in Fig.~\ref{CEF-phonon} (c) and show the field-evolution of the shoulder, disappearing quickly with increasing field strength. The normalization procedure distorts the original Lorentzian lineshape and introduces a kink around 61.5 meV manifesting as a vertical yellow strip between two blueish areas. (Fig.\ref{CEF-phonon}, b). This artifact is also observed in the experimental data (Fig.\ref{exp}, right panels) and 

stems from the considerable linewidth, which is larger than the field-induced shift of the peak position.  Assuming the phonon energy just below the CEF level, we obtain a field-evolution that is in great agreement with our data.  We repeated the simulation for a phonon energy that lies above the CEF level and get a completely different result, i.e., the shoulder would appear on the high energy side of the CEF transition (see Supplementary Materials for more details). We conclude, unlike what was reported in Gaudet et al. \cite{Gaudet}, that the phonon energy has to be lower than the CEF transition energy in order to get the observed response in IR spectroscopy.

Finally, it is worth noting that Hamiltonian eq.\ref{eq2} is just an approximation, allowing to qualitatively and semi-quantitatively describe our findings and explain why the vibronic shoulder shows up in our IR data for low magnetic fields only. 
For instance, the atomic displacements of the phonon mode have a complex influence to the crystal field of Ho$^{3+}$ ions and, hence, each quadrupole operator $\widehat{B}_{2}^q$ would have different coupling constant $g_q$. In addition, the presence of the vibronic coupling leads to some intensity on the fast moving CEF branch associated with E$_g$(3) transition, while only the transition to the lower E$_g$(3) doublet branch is magnetic-dipole allowed for $g_0=0$. In our experiment, the intensity of the higher lying branch might be obscured for low fields by the high intensity of the slow moving CEF branch, while for higher fields, its intensity is already disappeared as the phonon decouples from the CEF level as the field is increased (see Supplementary Materials for more details).

\section{Conclusions}

We have investigated the broadband magneto-optical response of HTO single crystals as a function of applied magnetic field. The weak magnetic-dipole excitations between CEF levels were revealed in the far-IR reflection signal on top of the strong electric-dipole phonon excitations. 
We model our magneto-IR spectra using the crystal-field Hamiltonian and a Zeeman term, leading to very good agreement with experimental observations.

Our results unambiguously determine the sign of the $B^{3}_{4} $ and $B^{3}_{6}$ coefficients, which is impossible in zero field measurements. Additionally, our spectroscopic data also clearly shows the presence of splitting of the E$_g$(3) CEF level at zero field, which we associate with a vibronic state. This vibronic state only appears at low field as its intensity quickly diminishes in applied magnetic fields. Modeling of the observed splitting shows that the phonon resides at slightly lower energy compared to the CEF level that it couples to, which is in contrast with previously published INS results~\cite{Gaudet}.

\section{Acknowledgements}
C.B. acknowledges support from the National Research Foundation, under grant NSF DMR-1847887. A portion of this work was performed at the National High Magnetic Field Laboratory, which is supported by National Science Foundation Cooperative Agreement No. DMR-1157490, No. DMR-1644779, and the State of Florida. H.D.Z acknowledges support from the NHMFL Visiting Scientist Program, which is supported by NSF Cooperative Agreement No. DMR-1157490 and the State of Florida. 

M.O and N.A. contributed equally to this work.
\typeout{} 
\bibliography{main}
\widetext
\begin{center}
\textbf{\large \textit{Supplemental Material:}Magnetic field tuning of crystal field levels and vibronic states in Spin-ice Ho$_2$Ti$_2$O$_7$ observed in far-infrared reflectometry}
\end{center}
\setcounter{equation}{0}
\setcounter{figure}{0}
\setcounter{table}{0}
\makeatletter
\renewcommand{\theequation}{S\arabic{equation}}
\renewcommand{\thefigure}{S\arabic{figure}}
\renewcommand{\bibnumfmt}[1]{[S#1]}
\renewcommand{\citenumfont}[1]{S#1}
\maketitle

\subsection*{Characterization of CEF levels and determination of optical phonons using absorption, Raman and IR spectroscopy}\label{phonons}
HTO has a cubic structure (lattice parameter 10.1~\AA), crystallizing in Fd$\bar{3}$m space group with eight formula units in a unit cell. The eight-coordinated Ho$^{3+}$ ions are located at \textit{16c} sites, whereas six-coordinated Ti$^{4+}$ ions are located at \textit{16d} sites, as shown in panel a) of Fig.~\ref{HTO}, both forming separate networks of corner sharing tetrahedra. The oxygen anions of one kind occupy \textit{48f} sites coordinating with two Ho$^{3+}$ and two Ti$^{4+}$ ions, whereas the oxygen anions of the other kind occupy \textit{8a} sites that are tetrahedrally coordinated with four Ho$^{3+}$ ions, also shown in panel a) and b) of Fig.~\ref{HTO} \cite{GardnerJ}.

Due to strong spin-orbit coupling in Ho$^{3+}$ ions, the 4$f^{10}$ energy level splits into several spin-orbit multiplets, $^5I_8$ being the ground state. Furthermore, extended 4\textit{f} orbitals result in strong orbital overlap with the surrounding oxygen atoms, leading to substantial crystal electric field effects. The symmetry of the CEF Hamiltonian partially lifts the 17-fold degeneracy of the J=8 state into six doublets and five singlets with a dominant $\left|m_{j}=\pm8\right\rangle$ ground-state doublet (see main text for CEF-level scheme), resulting in a strong local Ising anisotropy \cite{Rosenkranz,Tomasello}.

Based on lattice parameters, atomic Wyckoff positions, and lattice symmetry as shown in Fig.~\ref{HTO}b), the entire set of vibrational degrees of freedom is expressed in terms of the following irreducible point group representation at the center of the Brillouin zone,
\textbf{
\begin{equation}
\begin{aligned} 
\Gamma_{3N}= {} & \textcolor{red}{1A_{1g}+1E_{g}+4F_{2g}} \textcolor{blue}{+ 8F_{1u}} \textcolor{green}{+ 2F_{1g} +3A_{2u} + 3E_{u}}\\\textcolor{green}{ + 4F_{2u}} 
\end{aligned}
\end{equation}}
\noindent Here, \textit{N} denotes the total number of atoms in the primitive cell, which is 22 (4 Ho, 4 Ti, 12 \textit{f}-type and 2 \textit{a}-type O), as shown in panel b) of  Fig.~\ref{HTO}. All modes in red color represent Raman active modes (total 6 modes), while all in blue are infrared active modes (total 8 modes including one acoustic mode). The rest of the 12 modes in green are optically inactive  modes. There have been several experimental and first-principle studies on vibrational properties of HTO, which have tabulated all optically active and inactive phonon modes in the system. \cite{Lummen,Mkaczka,KUMAR,Ruminy,Kushwaha} 
In order to determine whether phonons play a role in the observed magnetic field dependence of the IR spectra of our HTO crystals, we have determined the optically active vibrational modes using Raman and IR spectroscopy.

Room temperature polarized Raman spectra were measured using a Horiba JY LabRam HR800 Raman spectrograph in the back-scattered geometry, supplying excitation wavelengths at 785 nm, 633 nm, 514 nm, 488 nm, 458 nm and 364 nm. LabRam HR800 was equipped with 600 and 1800 lines/mm gratings, providing a resolution of about 2--3 cm$^{-1}$ in the measurement region. The grating stabilized diode laser emitting 785 nm laser excitation was operated at 80 mW (15 mW at the sample), whereas the Melles-Griot 633 nm Helium-Neon laser was operated at 17 mW output power (6 mW at the sample). A coherent I-308 argon ion laser system allowing the Raman experiment at several wavelengths (514 nm, 488nm, 458nm, and 364 nm) was operated at about 20--30 mW of average power output. %While the 633 nm laser line has a strong overlapping electronic Raman signature, the 458 nm shows both vibrational and electronic features in the Raman spectra.   
\begin{figure}
    \centering
    \includegraphics[width = 0.9\textwidth]{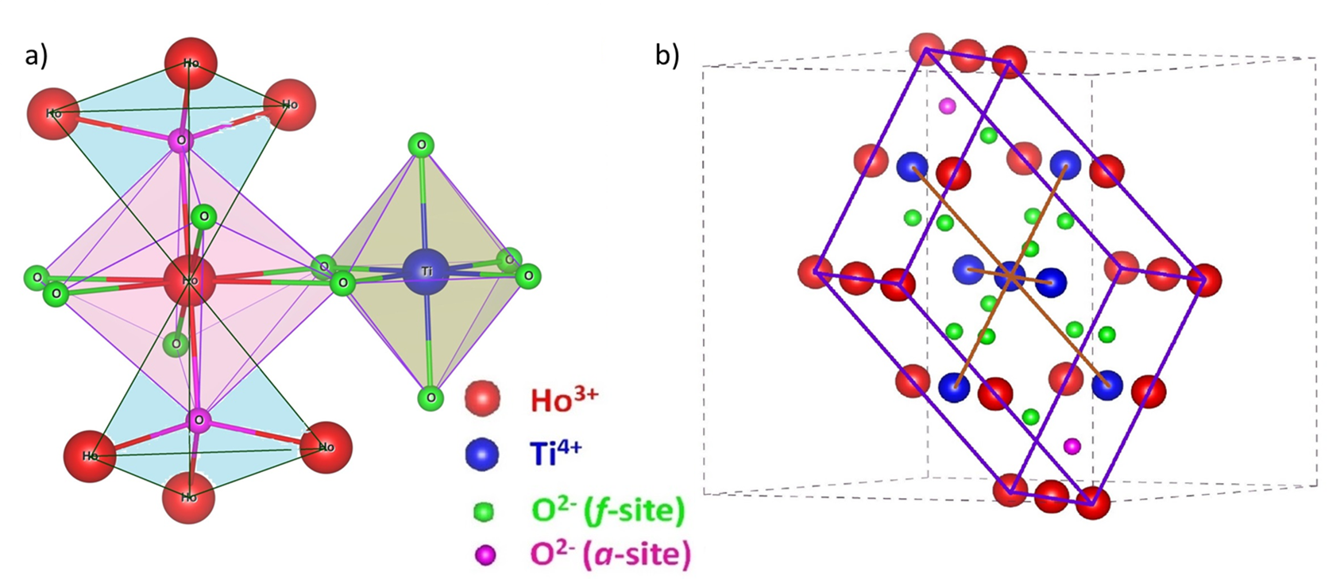}
    \caption{a) Crystal structure of HTO showing a) distorted cubic and octahedral crystal field around Ho$^{3+}$ and Ti$^{4+}$ ions respectively. b) Primitive unit cell showing interpenetrating FCC unit cells formed by both Ho$^{3+}$ and Ti$^{4+}$ ions. The primitive cell contains 4 Ho$^{3+}$, 4 Ti$^{4+}$, 12 \textit{f}-type and 2 \textit{a}-type O$^{2-}$ ions. }
    \label{HTO}
\end{figure}

\begin{figure}[h!]
\centering
\includegraphics[width =0.48\textwidth]{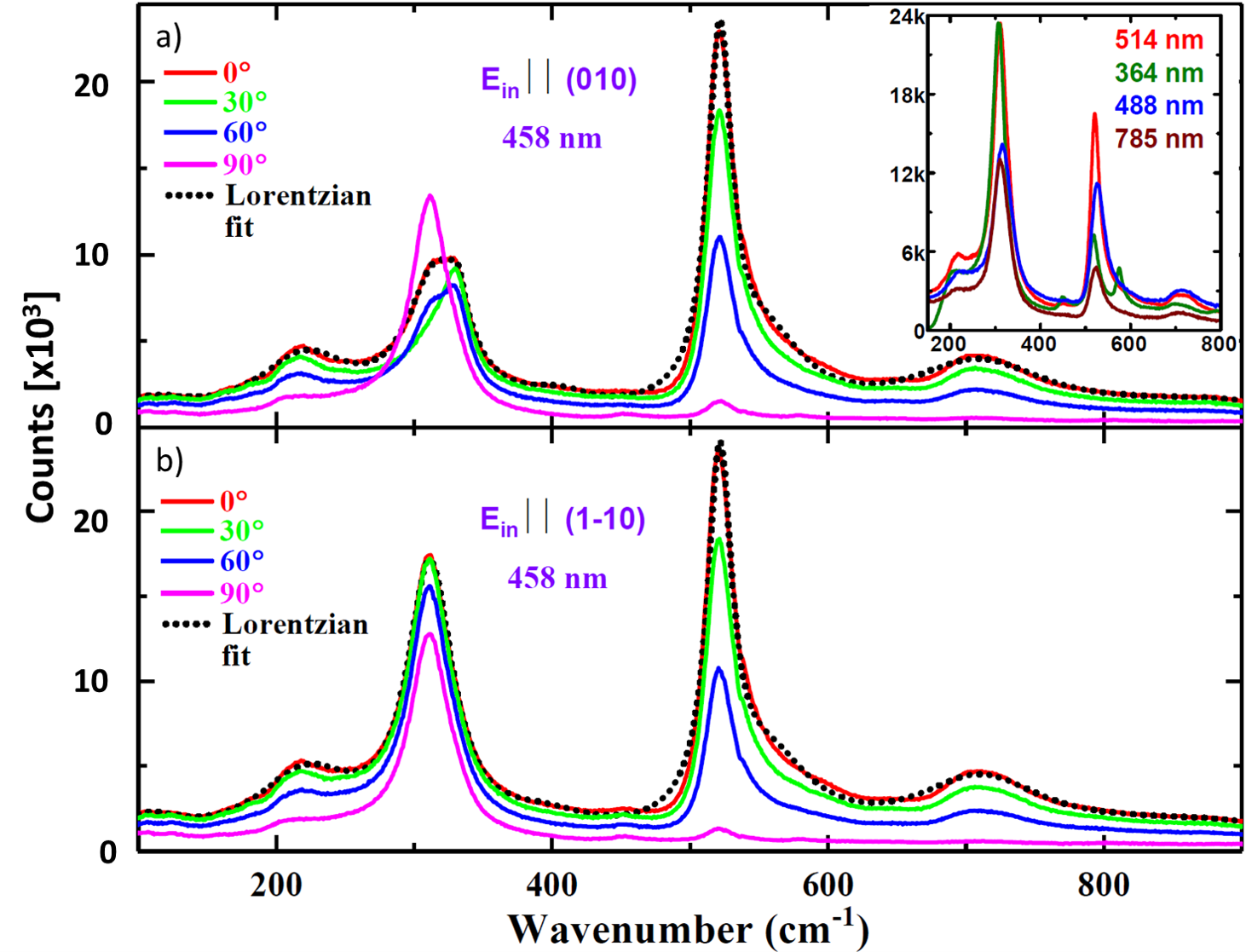}
\caption{\label{Raman}(Color online) Polarized Raman spectra on an oriented HTO single crystal using 458~nm laser line in back-scattered geometry. a) $\vec{E}_{in}||$ [010] and b) $\vec{E}_{in}||$ [1$\bar{1}$0] (Analyzer axis $\parallel \vec{E}_{in}$ = 0$^{\circ}$, Analyzer axis $\perp \vec{E}_{in}$ = 90$^{\circ}$). Inset: Raman spectra using laser lines with $\lambda$ = 364, 488, 514, and 785~nm. The phonon mode frequencies obtained from Lorentzian multipeak fit are shown in Table \ref{Table 1}.}
\end{figure}

Room temperature polarized Raman spectra were performed on HTO crystals in back-scattered geometry for several polarizer-analyzer configurations and using various laser lines (Fig. \ref{Raman}). Panel a) shows spectra with $\vec{E}_{in}$ parallel to [010] axis while panel b) shows spectra with $\vec{E}_{in}$ parallel to [1$\bar{1}$0] axis. The analyzer is rotated in 30$^{\circ}$ steps between 0$^{\circ}$--90$^{\circ}$, where the 0$^{\circ}$ spectrum represents the configuration in which the polarizer and analyzer have parallel transmission axes. Spectra have been fitted with a Lorentzian model using HORIBA Scientific LabSpec 6 and the 0$^{\circ}$ fitted curve is included for both measurement configurations. Phonons were observed at 220 cm$^{-1}$ (F$_{2g}$),  310 cm$^{-1}$ (F$_{2g}$),  330 cm$^{-1}$ (E$_{g}$), 520 cm$^{-1}$ (A$_{1g}$) and  570 cm$^{-1}$ (F$_{2g}$) (see Table \ref{Table 1}). A very weak feature is observed at 450 cm$^{-1}$ (with $\lambda$ = 364~nm), which based on theoretical work \cite{Ruminy}, could be associated with an F$_{2g}$ mode. There is a weak band around 700 cm$^{-1}$, which is observed at all excitation wavelengths and is consistent with previous observations \cite{Lummen}. Based on the symmetries of the Raman-active modes, the parallel and perpendicular polarization spectra should discern F$_{2g}$ modes from E$_{g}$ and A$_{1g}$ modes. Although the room-temperature spectra of HTO indicate towards somewhat relaxed phonon selection rules, the mode assignments are performed based on their angular dependence and are in agreement with other reported studies \cite{Lummen}. 

\begin{table}[h]
\begin{center}
%\bgroup
\setlength{\tabcolsep}{10pt} % Default value: 6pt
\renewcommand{\arraystretch}{1.5} % Default value: 1
\caption{\label{Table 1} Optical phonons as determined from this work, which are in agreement with other reports  \cite{Lummen,Ruminy,Bi}. For the infrared set, transverse [longitudinal] modes are provided for 5 K.}
\begin{tabular}{|c | c | c|} 
 \hline
 
Freq. (cm$^{-1}$) & Assignment & Type  \\
\hline
84 [91]& $^1$F$_{1u}$ &IR\\\hline
130 [138]& $^2$F$_{1u}$ &IR\\\hline
220&  F$_{2g}$&Raman \\\hline
198 [252]& $^3$F$_{1u}$&IR\\\hline
265 [317]& $^4$F$_{1u}$ &IR\\\hline
310&  F$_{2g}$&Raman\\\hline
330&  E$_{g}$ &Raman\\\hline
371 [450]& $^5$F$_{1u}$&IR\\\hline
456 [535]&$^6$F$_{1u}$ &IR\\\hline
520& A$_{1g}$ &Raman\\\hline
550 [612]& $^7$F$_{1u}$&IR\\\hline
570& F$_{2g}$ &Raman\\\hline
613 [747]& F$_{1u}$ &IR\\\hline
\hline

\end{tabular}
%\egroup
\end{center}
\end{table}

Next, we focus on the infrared active phonons in HTO as determined from the IR spectrum shown in Fig. 1 of the main text. 
Table \ref{Table 1} lists all the transverse and longitudinal modes used to calculate the reflectance curve shown in the main text. The observed modes are in great agreement with previously reported values. Something that can be noted is that none of these optically active modes correspond to the energies at which we observe magnetic-field-dependent IR spectral features. 

\begin{figure}[h!]
    \centering
    \includegraphics[width = 0.95\textwidth]{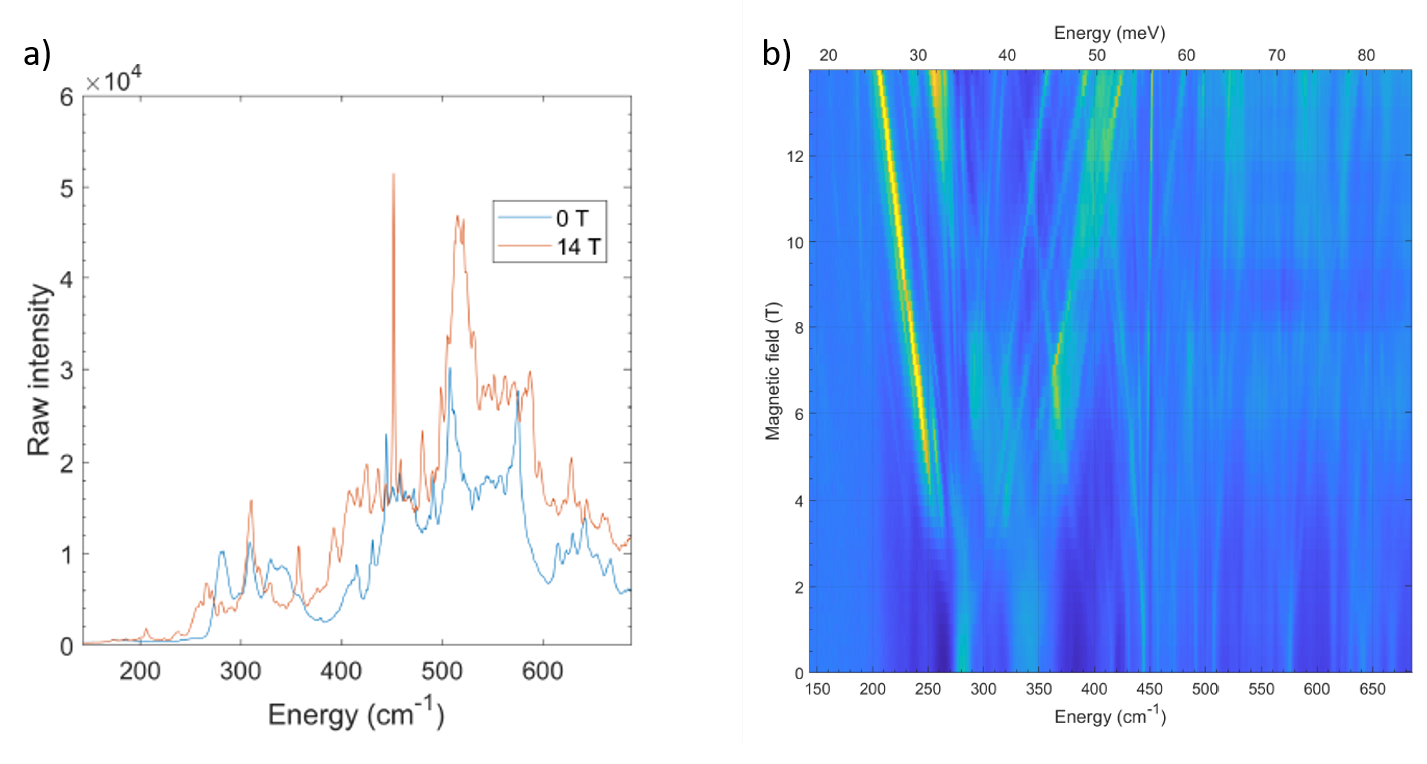}
    \caption{ a) Resonant Raman spectra on HTO using $\lambda$ = 532~nm taken at 5~K and B = 0 and 14~T focusing on the spectral range containing phonons and $^5$I$_8$ CEF transitions. b) Color map of normalized (with the average spectrum taken at all fields) Raman spectra (532~nm) collected at 5~K as a function of magnetic field. } 
    \label{Magneto-Raman}
\end{figure}

Field dependent Raman spectroscopy was performed in a back-scattering Faraday geometry in B up to 10 T ($B||[111]$) at several temperature points. 
We used a 532~nm excitation to look for possible changes in the Raman phonon modes, shifts in CEF levels, or changes in the relaxation mechanism of the excited carriers. This excitation wavelength is still resonant with CEF transitions, leading to a complicated response in the 100~-~700~cm$^{-1}$ range. Fig. \ref{Magneto-Raman} a) shows spectra taken at 0~T and 14~T, with Fig. \ref{Magneto-Raman}b) showing the field dependence of the Raman spectra normalized with respect to the average spectrum (taken over all field points).The color map clearly shows many field-dependent features of both phononic and CEF nature, making the Raman spectra hard to analyze.    

\subsection*{Modeling of far-Infrared Spectra in applied field}
\begin{figure}[h]
\centering
\includegraphics[width =0.9\textwidth]{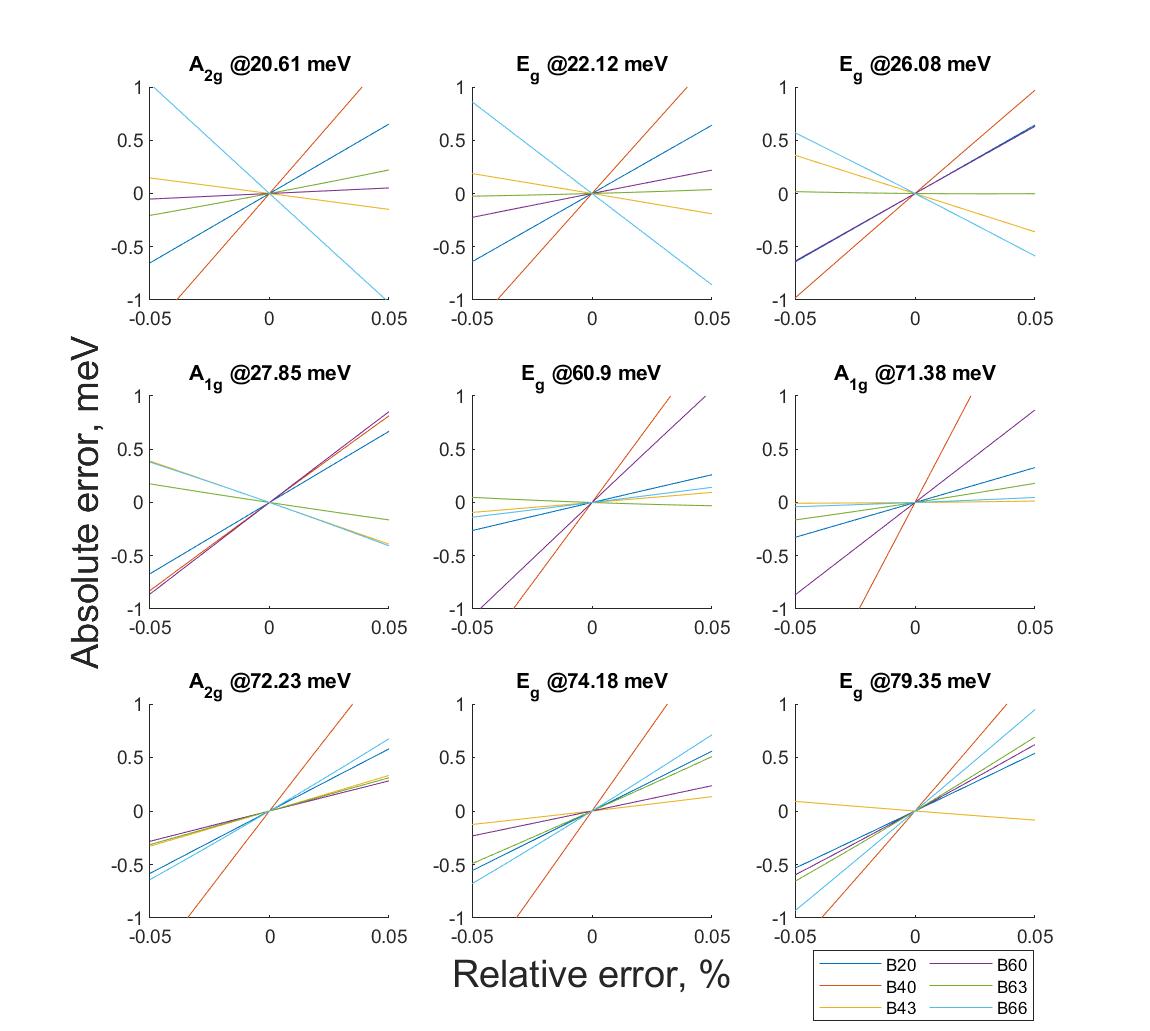}
\caption{\label{accuracy}(Color online) Variation in CEF transition energy as a function of changes in the CEF parameters.}
\end{figure}
We employed the EasySpin software package \cite{Easyspin} to solve the crystal field Hamiltonian and find the positions and intensities of the magnetic-dipole excited transitions between CEF levels. 
In the main text, zero-field CEF parameters previously reported by others \cite{Bertin,Tomasello,Rosenkranz,Gaudet,Ruminy} are compared. We show the simulated IR spectra (in zero applied field) for each of the CEF parameter sets presented in the main text. A small amount of artificial white noise and convolution with Lorentzian broadening were added to the spectrum. Noticing some mismatch between reported CEF parameter values and measured CEF level energies, we use the powerful fitting toolbox of EasySpin to determine optimized CEF parameter values. EasySpin allowed us to include both the energy positions as well as intensities into the fit, where six $B^{q}_{k}$ were the only fitting parameters (the magnetic field was fixed at zero). The best fit values for $B^{q}_{k}$ are presented in the main text. We use these optimized zero-field crystal field parameters to model our field dependent data. Fig. \ref{accuracy} shows how much each of the CEF transition energies will  be affected by a 5$\%$ change in each of the CEF parameter values ($B^{q}_{k}$). On average, changes of about 5$\%$ can lead to shifts of up to 1 meV in the transition energies.  
\begin{table}[h]
\begin{center}
%\bgroup
\setlength{\tabcolsep}{10pt} % Default value: 6pt
\renewcommand{\arraystretch}{1.5} % Default value: 1
\caption{\label{wave} Tabulated wavefunctions of the crystal field states of HTO obtained using each of the CEF parameter sets presented in the main text in Table I. For each of the $m_J$ values of the ground state multiplet, with only the first member of the doublet shown.  }
\begin{tabular}{|c|c | c | c|c|c|} 
 \hline
 
$m_J$&Tomasello \cite{Tomasello}& Bertin\cite{Bertin}& Ruminy \cite{Ruminy}& Gaudet\cite{Gaudet} & This Work \\
\hline
8 &       0         &  0        &   0&           0 &          0\\\hline
 7  &-0.0073&      0.0055&      0.0069&      0.0068&      0.0065\\\hline
 6    &    0    &       0      &     0  &         0      &     0\\\hline
 5   &     0    &       0   &        0   &        0      &     0\\\hline
 4 &  0.0573    &  0.0316      &0.0541  &    0.0518    &  0.0491\\\hline
 3   &     0    &       0     &      0   &        0      &     0\\\hline
 2     &   0      &     0    &       0  &         0     &      0\\\hline
 1&  -0.0744    &  0.0714 &     0.0731  &    0.0704  &    0.0705\\\hline
 0      &  0     &      0    &      0    &       0       &    0\\\hline
-1   &     0     &      0     &      0  &         0    &       0\\\hline
-2 &  0.0766  &    0.0137   &   0.0747  &    0.0727  &    0.0639\\\hline
-3  &      0  &         0    &       0   &        0      &     0\\\hline
-4  &      0    &       0    &       0   &        0       &    0\\\hline
-5 & -0.1536 &     0.1907     & 0.1542   &   0.1509  &    0.1562\\\hline
-6 &       0    &       0      &     0    &       0     &      0\\\hline
-7  &      0      &     0       &    0    &       0     &      0\\\hline
-8  & 0.9806   &   0.9784   &    0.981   &    0.982   &   0.9819\\\hline

\hline

\end{tabular}
%\egroup
\end{center}
\end{table}

To further compare our optimized CEF parameters to previously reported values, we determined and tabulated the wavefunctions for each of the $m_J$ values of the ground state multiplet for all parameters presented in the main manuscript (see Table \ref{wave}). Quantum corrections to the classical model \cite{RauGingras,Ruminy} stem from spectral content of sub-leading components of the wave function. As expected, we find the spectral content to be predominantly $|\pm8>$ with the sub-leading components of the wave function comparable to those presented by others \cite{Tomasello,Ruminy,Bertin}.

\begin{figure}[h!]
%\vspace{-0.41cm}
\includegraphics[width=0.4\textwidth]{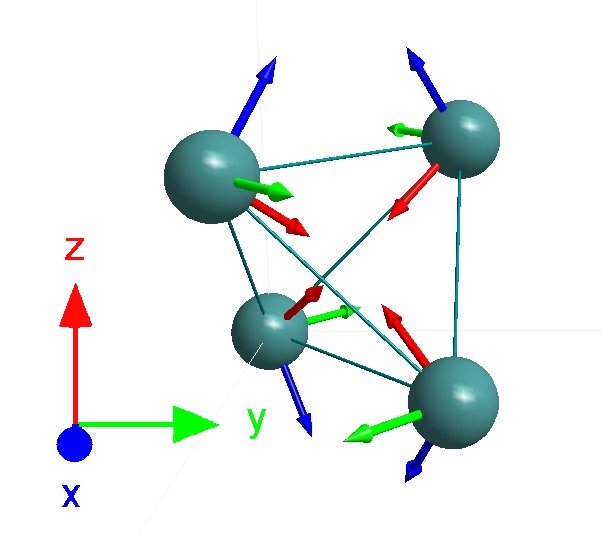}
\caption{\label{euler} (color online)  (Color online) One of the corner-sharing Ho$^{3+}$ (green spheres) tetrahedra is shown, the x$_i$ (blue arrow), y$_i$ (green arrow), and z$_i$ (red arrow) indicate the site specific reference frame for each of the sites, and the X, Y, Z directions indicate the laboratory reference frame.} 
\vspace{+0.3cm}
\end{figure}

Applying magnetic field we notice that HTO has four non-equivalent sites for the Ho$^{3+}$ magnetic ions that reside on the vertices of the corner-sharing tetrahedra. The CEF Hamiltonian shown in the main text is written in the local coordinate frame of each site, which is related to the global coordinate frame via rotation by the following paired Euler angles (see Fig. \ref{euler}).

\begin{equation*}
\begin{array}{l}\label{2}
\alpha_1 =\frac{\pi}{4}
; \quad
\beta_{1} =arccos\frac{1}{\sqrt{3}} ;  \gamma = 0;\\
\alpha_{2}=\alpha_{1};   \quad  \beta_{2}=\pi-\beta_1 ; \gamma = \pi ;\\
\alpha_{3}=\pi+\alpha_{1};  \quad  \beta_{3}=\pi-\beta_1 ; \gamma =\pi;\\
\alpha_{4}=- \alpha_{1}; \quad  \beta_{4}=\beta_1 ; \gamma=0;\\
\end{array}
\end{equation*}

\begin{figure}[h]
\centering
\includegraphics[width =0.65\textwidth]{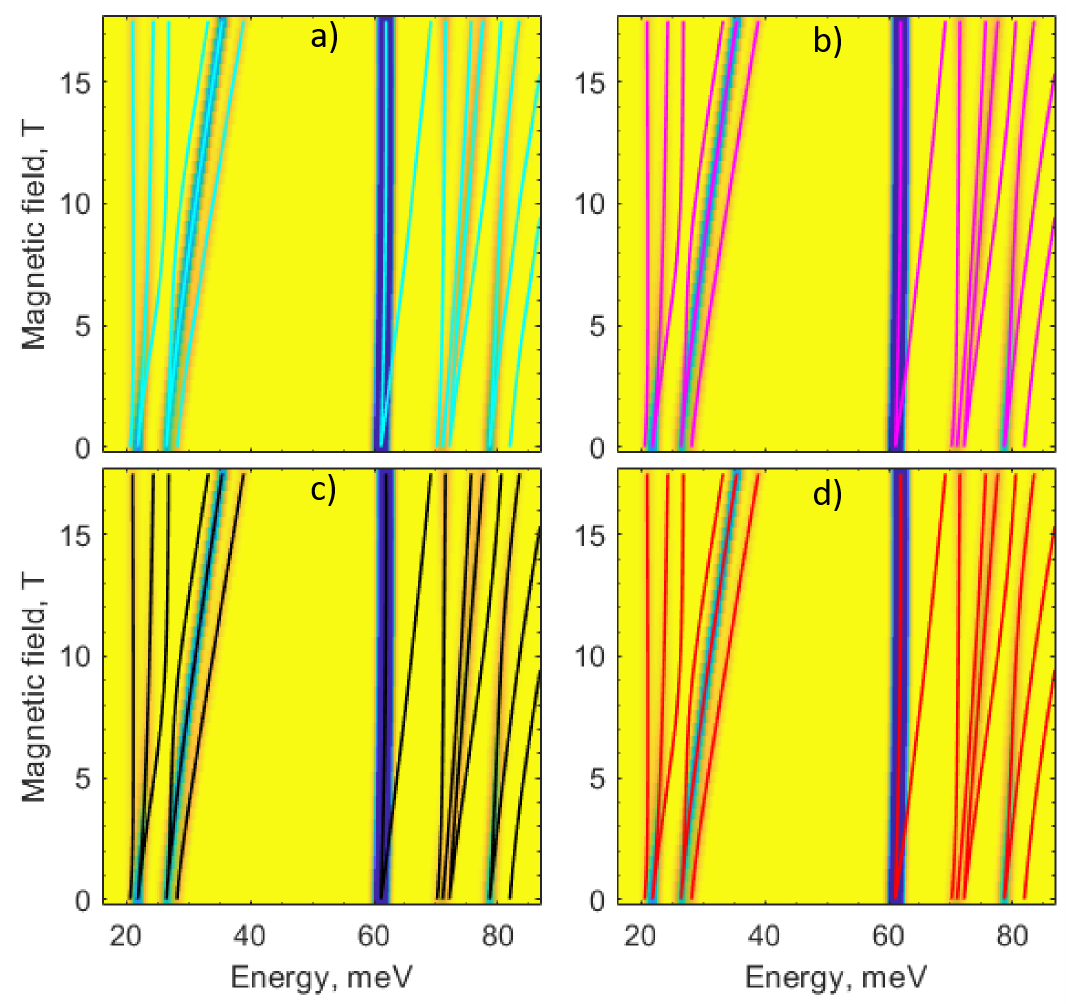}
\caption{\label{sites}(Color online) The Ho$^{3+}$ ions reside on cornersharing tetrahedra. The four sites within one tetrahedron (see Fig. 3 in the main text) are not equivalent. Panels a) - d) show the calculated intensity of the CEF transitions for each of these sites as a function of applied magnetic field. The solid lines show the energy difference between high-energy and lowest CEF energy levels and their splitting in the applied magnetic field.}
\end{figure}

Hence, we start by simulating the IR spectra as a function of applied magnetic field for each of the individual sites. In Fig. \ref{sites} we show the calculated intensity for each of the inequivalent Ho$^{3+}$ sites as function of applied magnetic field (based on the Bertin\cite{Bertin} CEF parameters). The solid lines show the energy difference between high-energy and lowest CEF energy levels and their splitting in the applied magnetic field. This shows that there are no significant differences between the sites when it comes to their contributions to the IR spectroscopic response in applied magnetic field. Hence, in the following we simply average the contributions from the sites to generate color maps of calculated IR intensity. 

%To show that the parameters taken from Bertin et al \cite{Bertin} provide a better basis for our model, we start by presenting the calculated energies based on the $B^{q}_{k}$ coefficients from Tomasello et al. \cite{Tomasello}, which similar to what we present in the main text, also does not show a good agreement with the zero-field experimental data as reported in the most recent experimental work on neutron scattering ref.\cite{Gaudet} (see Fig. \ref{Tom-zero}). 

\begin{figure*}%\vspace{-1in}
\includegraphics[ width=\textwidth]{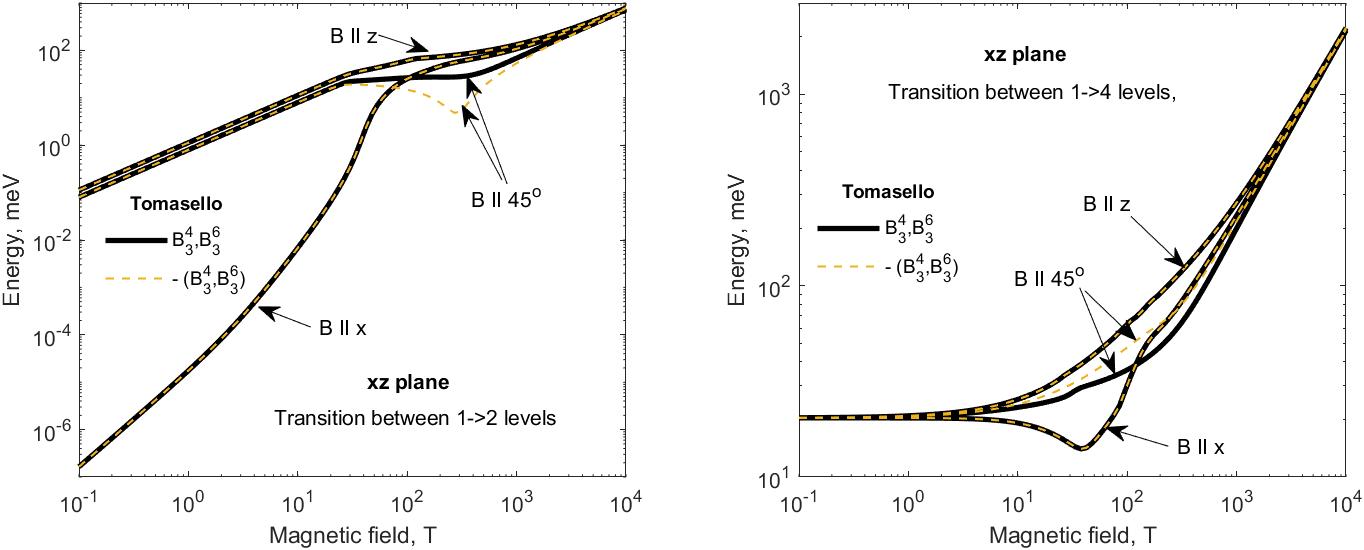}
\caption{\label{xz}(color online) (Left) Calculated field dependence of the transition energy  between first and second CEF levels (E$_g(0)_-$ $\rightarrow$ E$_g(0)_+$). The magnetic field is applied in the xz plane of the Ho$^{3+}$ local coordinate frame. The black line corresponds to the calculations using CEF parameters taken from Tomasello’s article~\cite{Tomasello}. The orange line is calculated for the ($B_4^3$, $B_6^3$) coefficients taken with reversed sign. The overlap between the lines is observed for magnetic field applied parallel to the $x$ and $z$ axes only. (Right) The same dependence is calculated for the transition between first and fourths CEF levels (E$_g(0)_-$ $\rightarrow$ E$_g(1)_-$).
}
\end{figure*}

\begin{figure*}%\vspace{-1in}
\includegraphics[ width=\textwidth]{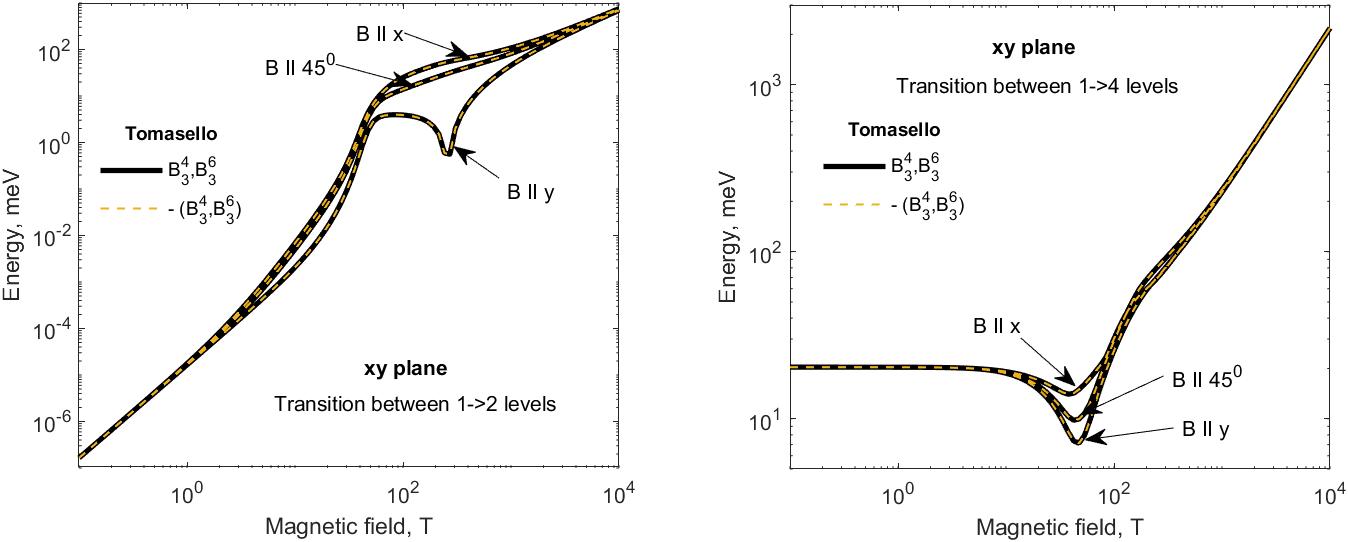}
\caption{\label{xy}  (color online) (Left) Calculated field dependence of the transition energy  between first and second CEF levels (E$_g(0)_-$ $\rightarrow$ E$_g(0)_+$). The magnetic field is applied in the xy plane of the Ho$^{3+}$ local coordinate frame. The black line corresponds to the calculations using CEF parameters taken from Tomasello’s article~\cite{Tomasello}. The orange line is calculated for the ($B_4^3$, $B_6^3$) coefficients taken with reversed sign. No difference after sign exchange is found for all field orientations. (Right) The same dependence is calculated for the transition between first and fourths CEF levels (E$_g(0)_-$ $\rightarrow$ E$_g(1)_-$) .}

\end{figure*}

\begin{figure}[h]
\centering
\includegraphics[width =0.7\textwidth]{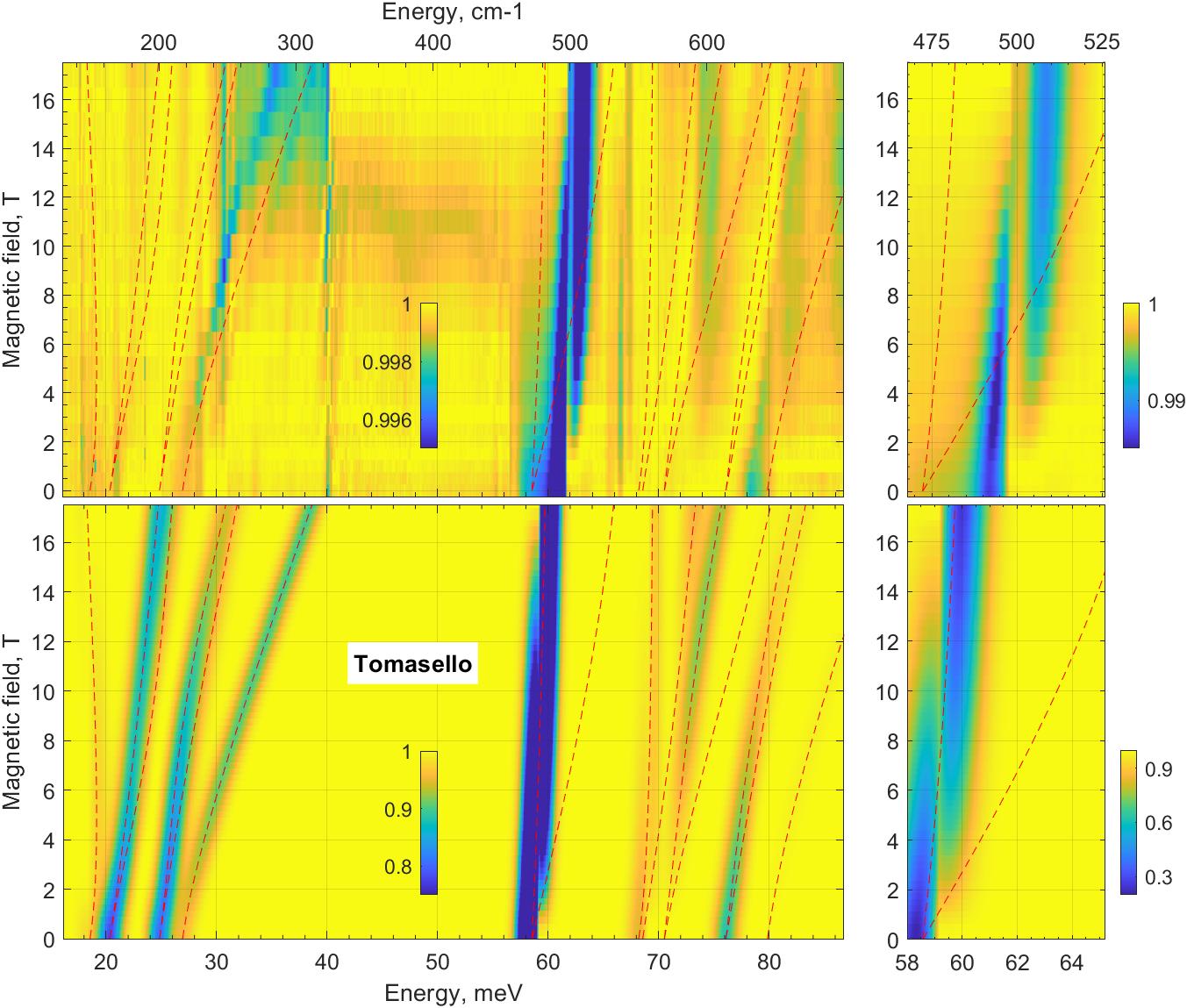}
\caption{\label{Tom-H}(Color online) (top) The same experimental data  presented in the Fig.3 of the main text. (Bottom) Simulations were using crystal field parameters presented by Tomasello et al. \cite{Tomasello}.}
\end{figure}

\begin{figure}[h]
\centering
\includegraphics[width =0.7\textwidth]{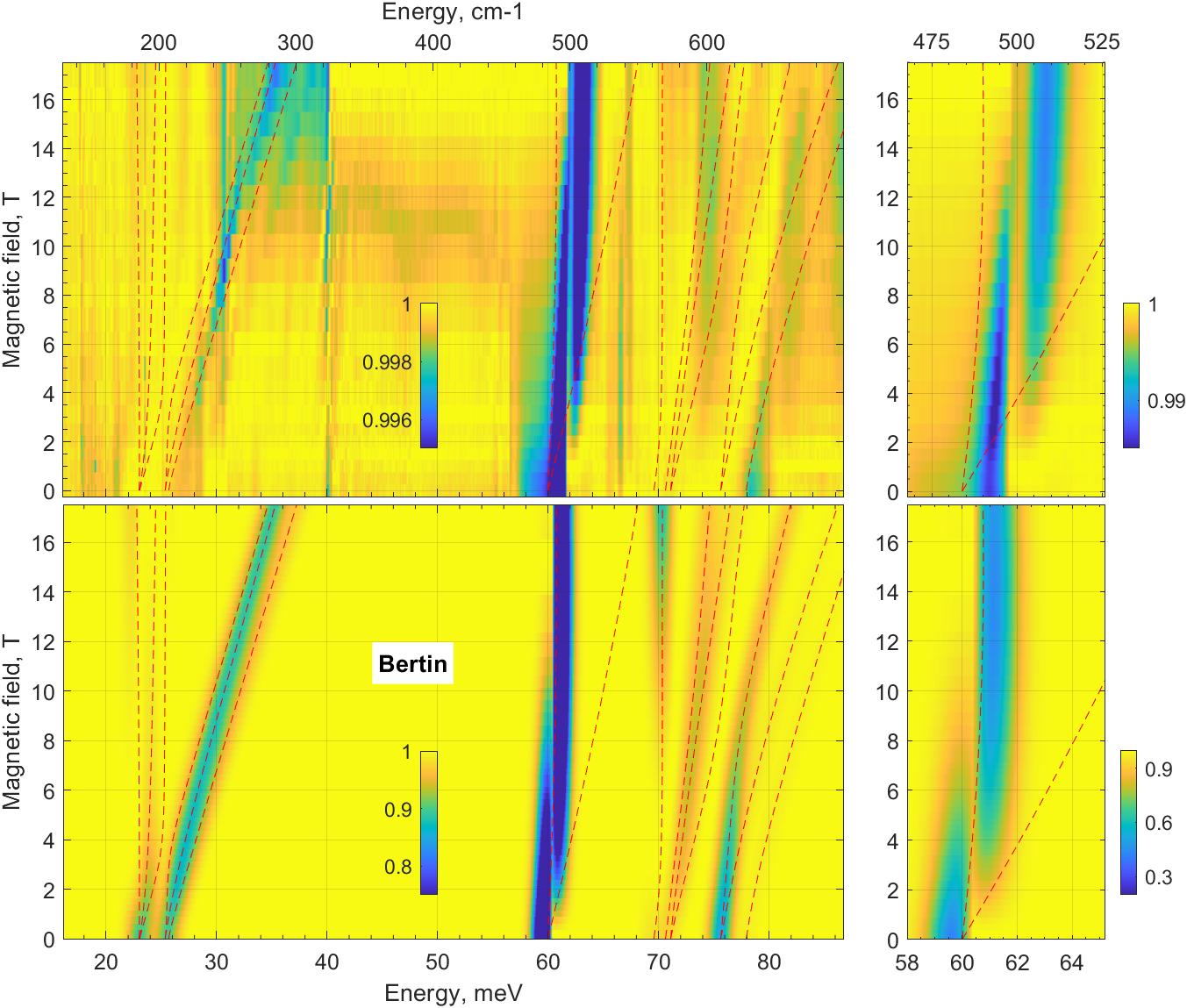}
\caption{\label{Bertin-H}(Color online) (top) The same experimental data presented in the Fig.3 of the main text. (Bottom) Simulations were using crystal field parameters presented by Bertin et al. \cite{Bertin}.}
\end{figure}

\begin{figure}[h]
\centering
\includegraphics[width =0.7\textwidth]{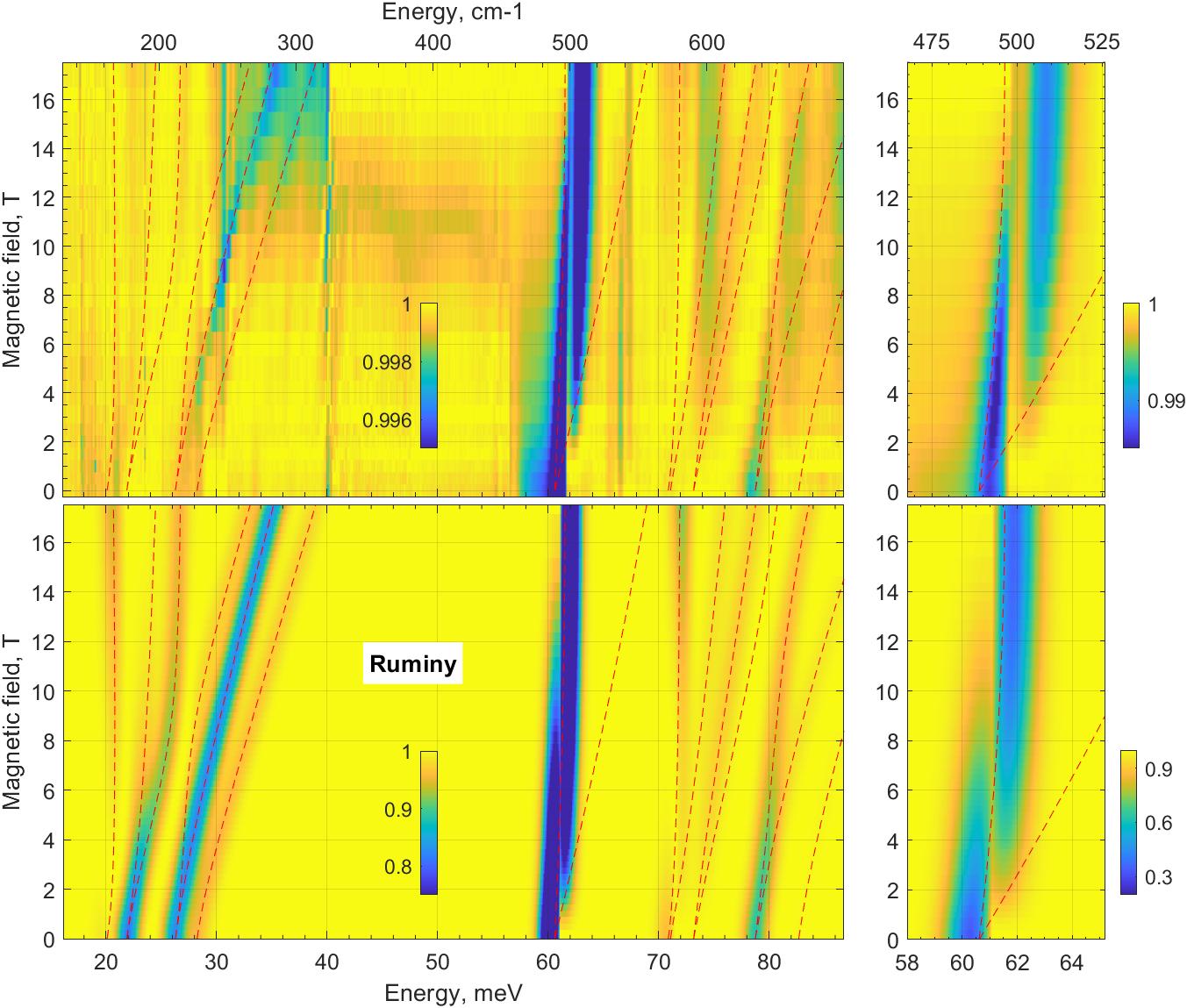}
\caption{\label{Ruminy-H}(Color online)  (top) The same experimental data  presented in the Fig.3 of the main text. (Bottom) Simulations were using crystal field parameters presented by Ruminy et al. (LS-coupling scheme) \cite{Ruminy}.}
\end{figure}

\begin{figure}[h]
\centering
\includegraphics[width =0.7\textwidth]{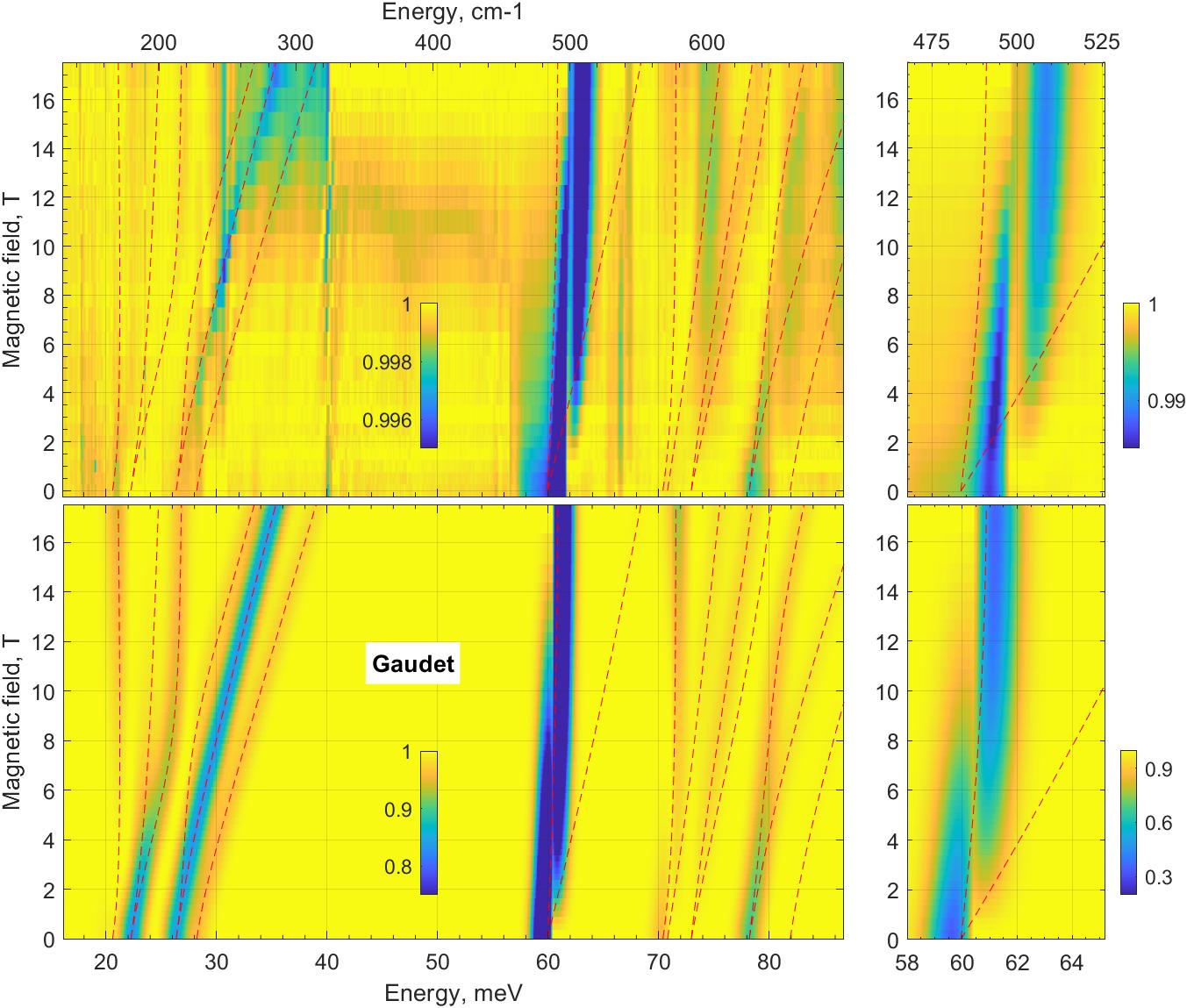}
\caption{\label{Gaudet-H}(Color online) (top) The same experimental data  presented in the Fig.3 of the main text. (Bottom) Simulations were using crystal field parameters presented  by Gaudet et al. \cite{Gaudet}.
}
\end{figure}

Similar to what we present in the main text, we used the powerful fitting toolbox of EasySpin to model the field dependence of IR spectroscopic features associated with the CEF transitions for each previously reported $B^{q}_{k}$ parameter set (see Table I in the main text). Figs.~ \ref{Tom-H},~\ref{Gaudet-H},~\ref{Ruminy-H}, and \ref{Bertin-H} show color maps of the calculated intensity of CEF transitions as a function of applied magnetic field (bottom panels). The solid red lines show the energy difference between the high-energy and the lowest CEF energy levels and their splitting in applied magnetic field.  We compare each of these calculated color maps to our measured spectroscopic response (top panels). The field dependence of the CEF levels based on the parameters presented by Tomasello \cite{Tomasello} and Bertin \cite{Bertin} show clear discrepancies for the field-evolution of the E$_g$(2) CEF level and both parameter sets underestimate the energy of the transition associated with E$_g$(3). The parameters presented by Ruminy \cite{Ruminy} and Gaudet \cite{Gaudet} describe the transitions observed in far-IR spectroscopy quite well, but both still underestimate the energy for the transition associated with E$_g$(3).  
We calculated the transition energies as a function of applied field strength and direction using the Hamiltonian described in the manuscript. We show the effect of a sign reversal of the B$^{3}_{4}$ and B$^{3}_{6}$ coefficients using the Tomasello coefficients. In Fig. \ref{xz} we plot the behavior of the transition between the ground state doublet and between the ground state and Eg(1) levels while the field direction is varied in the xz-plane. In Fig. \ref{xy} we plot the same transitions as a function of field, but now with the field direction varied within the xy-plane. What these figures show is that qualitatively the sign change in the coefficients only results in an actual change in the observed transition energies when the field is applied in very specific directions within the crystal electric field frame. When the field is applied anywhere in the xy-plane (this is what Tomasello et al. \cite{Tomasello} does), the sign changes do not affect the transition energies even in applied fields. This is also true for fields along the z-axis. Because we apply the field in the [001] direction of the global or lab frame (i.e., the field is along a $<$111$>$ in the local frame) in our experiments, we are able to distinguish the sign of the coefficients. Different CEF transitions are not affected in the same way, i.e., the field strength needed to be able to distinguish the signs is different for each of the transitions.

\begin{figure}[h]
\centering
\includegraphics[width =0.7\textwidth]{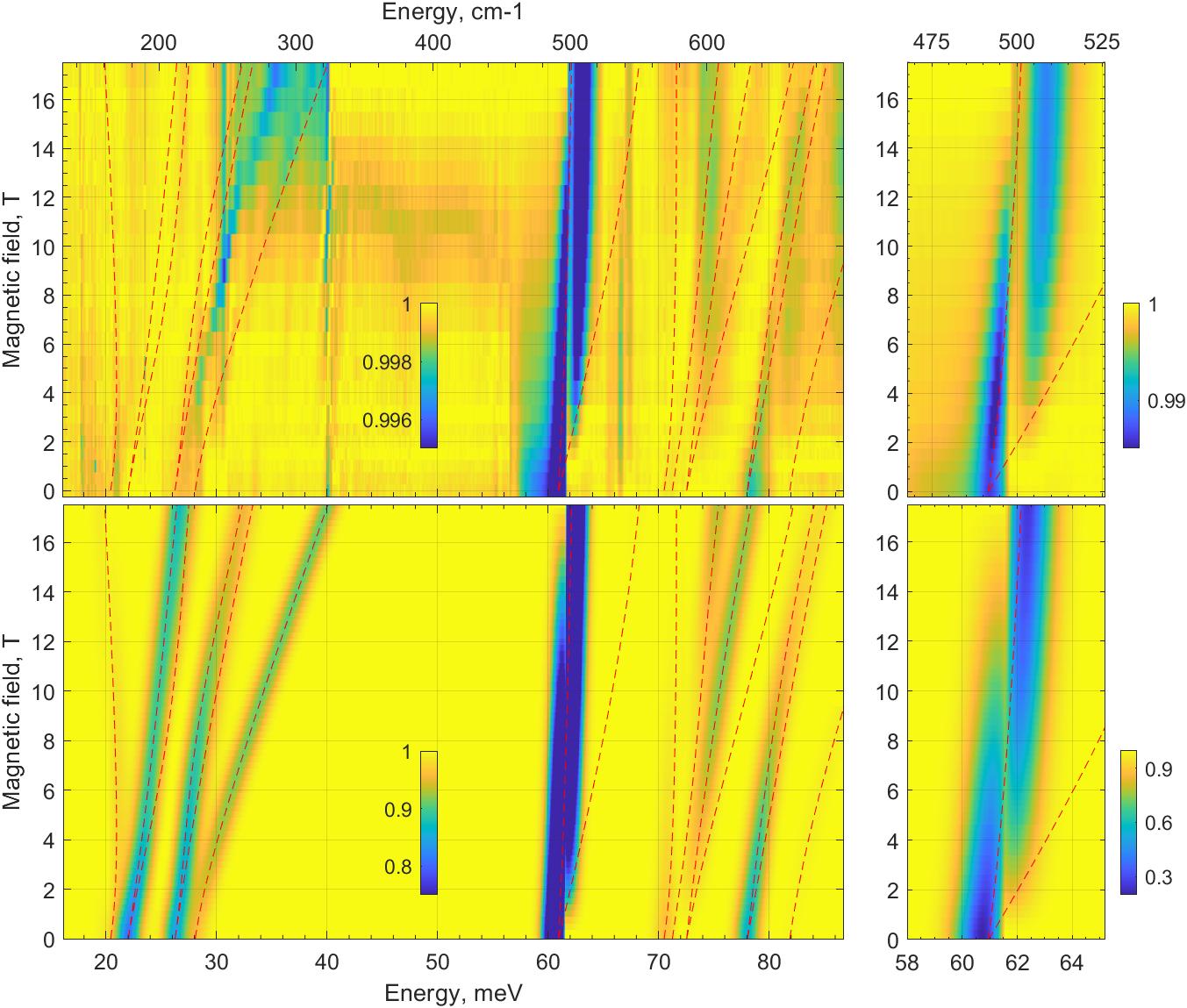}
\caption{\label{exp_sign}(Color online) (top) The same experimental data presented in the Fig.3 of the main text. (Bottom) Simulations were using the optimized crystal field parameters with  ($B_4^3$, $B_6^3$) coefficients taken with reversed signs. }
\end{figure}

\begin{figure}[h]
\centering
\includegraphics[width =0.5\textwidth]{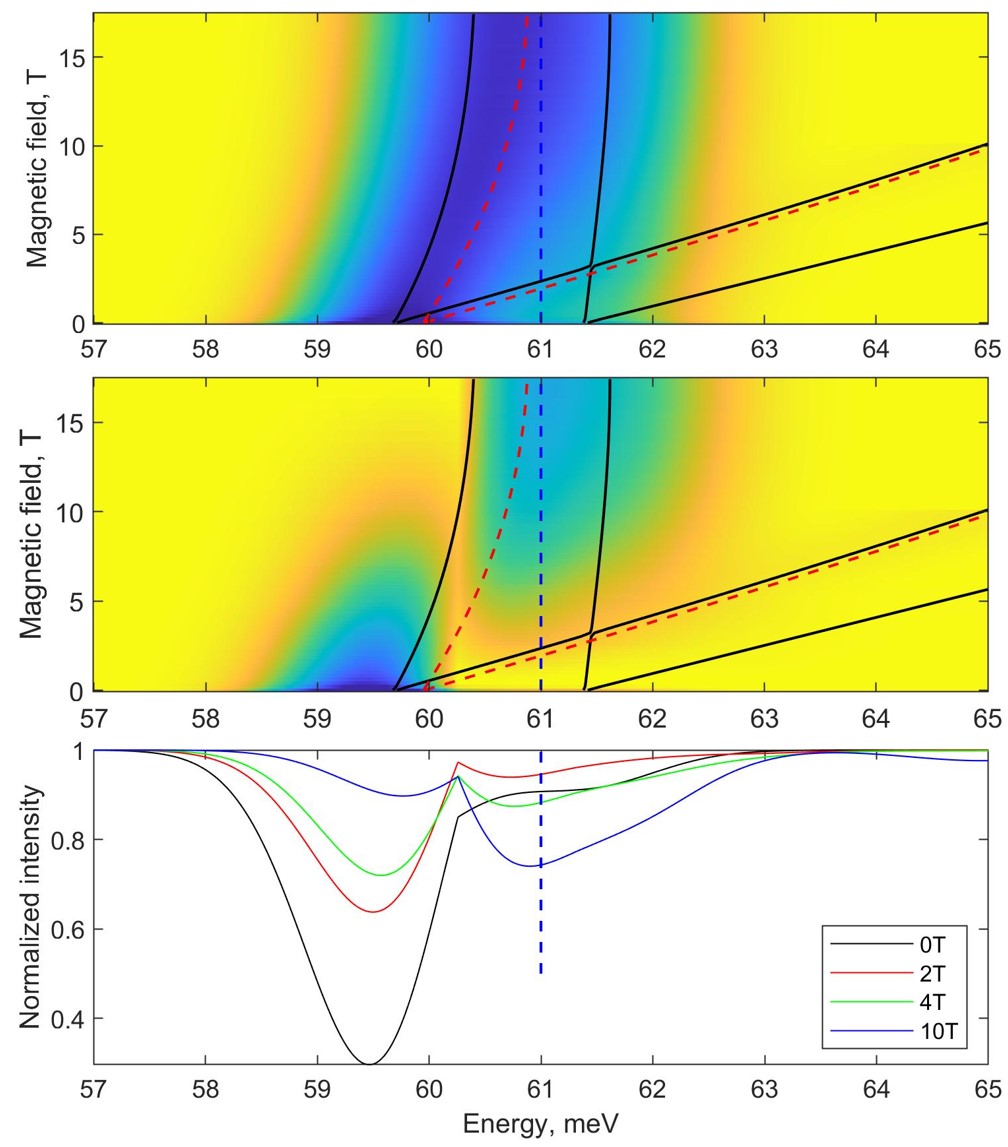}
\caption{\label{VIB-SM}(Color online) Top) Illustration of the excitation spectrum from ground to hybridized states as a function of the magnetic field.  The intensity of the transitions is calculated for $T=0$~K, $\hbar\omega=61$~meV, coupling constant $ g_0=0.016$  and Lorentizan linewidth of 1.6~meV. The dashed red and blue lines correspond to the  $E_g(3)$ CEF doublet and phonon mode, respectively. The black solid lines show energies of hybridized states. Middle) The same excitation spectrum but now normalized by a reference spectrum, calculated in the same way as was applied for the Fig.~3 in the main text.  Bottom) Profiles taken at various fields based on the middle panel.}
\end{figure}

We simulate the IR spectrum around 61~meV using the vibronic Hamiltonian defined in the main text, similar to previous reports, we do not take into account any splitting of the ground state due to the presence of the vibronic coupling. In the top panel of Fig. \ref{VIB-SM} we show the calculated IR absorption line associated with the E$_g$(3) CEF transition in the presence of phonon-CEF hybridization, with the phonon energy indicated by the dashed black line, i.e., at higher energy compared to E$_g$(3) at zero field. The red dashed lines are the CEF transitions and their field dependence based on the CEF Hamiltonian only. The colormap clearly indicates the presence of two well-resolved features around 61~meV. To compare to our measured data we apply the same normalization routine as before, which results in the color map in Fig. \ref{VIB-SM} (center panel). Profiles taken at B = 0, 2, 4 and 10 T result in the bottom panel and show the field-evolution of the split CEF level. By assuming the phonon energy to be just above 61~meV, i.e., above the CEF level, we obtain a field-evolution that is completely different from our observations in IR spectroscopy.  If the phonon is at higher energy compare to the CEF it couples to, the shoulder or resolved peak would appear on the high energy side of the CEF transition. This clearly shows that unlike what was reported in Gaudet et al. \cite{Gaudet}, that the phonon energy has to be lower than the CEF transition energy in order to get the observed response in IR spectroscopy.    

%\textcolor{blue}{Do we need figure \ref{B22} and the accompanying red text?}
\begin{figure*}%\vspace{-1in}
\includegraphics[ width=\textwidth]{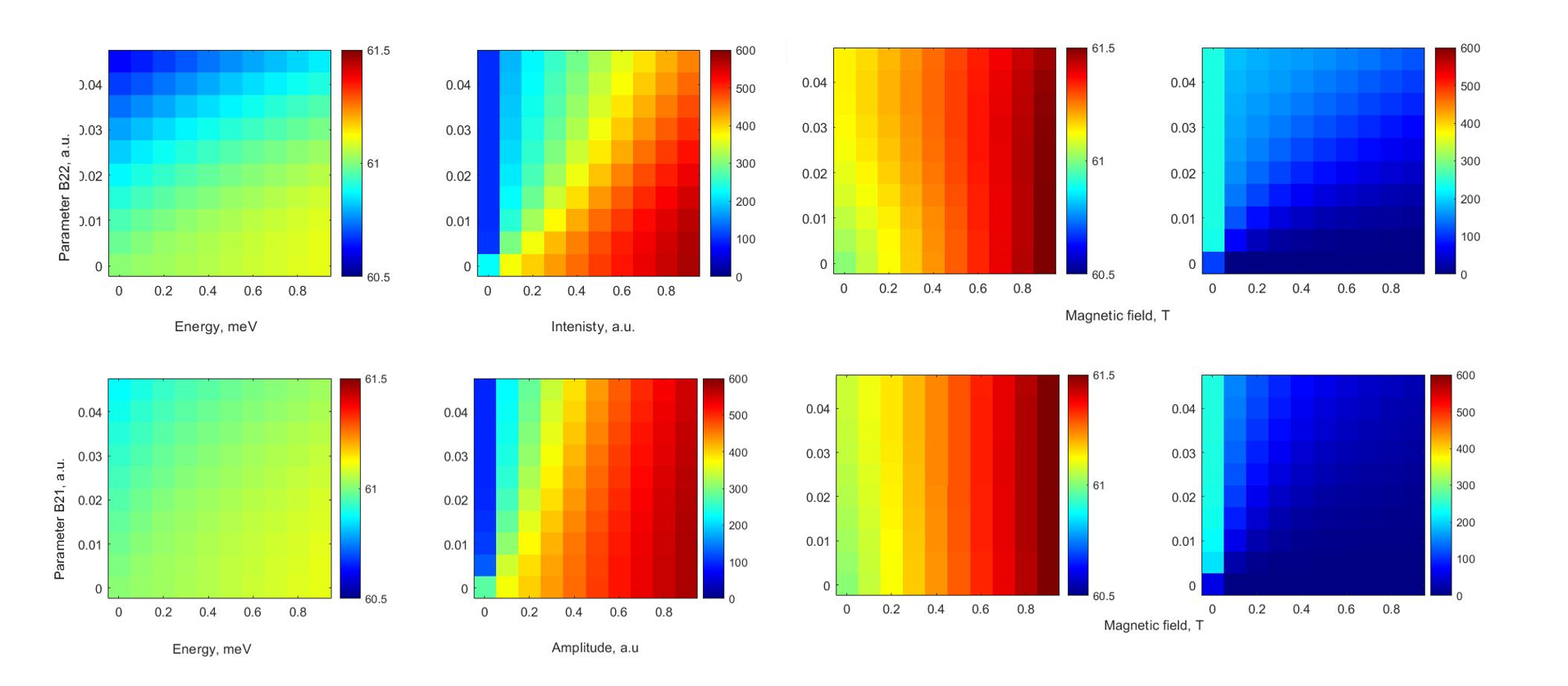}
\caption{\label{B22} (color online) Overview of how B22 and B12 coefficients affect the energy and intensity of the 60 meV CEF as a function of applied field. Left two panels: the slow moving CEF branch. right two panels: the faster moving CEF branch. }
\end{figure*}
It is worth noting that to model this CEF-phonon hybridization,  quadrupolar operators ($\hat{B}^q_2(\pm 1, \pm 2)$) become non-zero, leading to symmetry breaking of the CEF. When we model the IR intensity of the CEF transitions in the presence of this symmetry breaking, we find that in applied field, both CEF branches that originate from the E$_g$(3) CEF level have non-zero intensity in IR spectroscopy. The field dependence (up to 1~T) of both branches in IR transmission (transition energy and intensity) is presented in Fig. \ref{B22}. The lower energy branch shifts slowly to higher energy and increases in intensity with applied field. This branch has non-zero intensity in IR even if the quadrupolar operator coefficients are zero. When these operators are assigned larger coefficients this branch is shifted to slightly lower energy and the intensity rises more slowly in applied field. The second branch has no intensity in any applied field unless  the quadrupolar coefficients are given a nonzero value. The effect of each of the quadrupolar operators appears to be very similar.  The energy shifts faster as a function of applied field  while the IR intensity of this transition decreases quickly in increased magnetic field. While the vibronic coupling allows the fast moving branch to show intensity at low fields, this cannot be resolved in our measurements because of the presence of the much more intense transition (i.e., the lower energy CEF branch). 
\clearpage
\end{document}